\documentclass[a4paper,11pt]{article}
\pdfoutput=1 % if your are submitting a pdflatex (i.e. if you have
\usepackage{jinstpub} % for details on the use of the package, please
\usepackage{amsfonts}
\usepackage{booktabs}
\usepackage{lineno}
\usepackage{graphicx}
\usepackage{xcolor}

\newcommand{\pbfd}{PbF$_2\,$}

\newcommand{\temp}{$^{\circ}$C }

\usepackage[numbers,sort&compress]{natbib}
\begin{document}
\title{\boldmath Crilin: A CRystal calorImeter with Longitudinal InformatioN for a future Muon Collider}

\author[a]{S. Ceravolo,}
\author[b]{F. Colao,}
\author[c]{C. Curatolo,}
\author[a]{E. Di Meco,}
\author[a]{E. Diociaiuti,}
\author[d]{D. Lucchesi,}
\author[a]{D. Paesani,}
\author[e]{N. Pastrone,}
\author[f]{A. Saputi,}
\author[a]{I. Sarra\note{Corresponding author.},}
\author[d]{L. Sestini,}
\author[g]{D. Tagnani}

\affiliation[a]{INFN, Laboratori Nazionali di Frascati, Via Enrico Fermi 54, 00054 Frascati, Italy}
\affiliation[b]{Enea Frascati, Via Enrico Fermi 45, 00044 Frascati, Italy}
\affiliation[c]{INFN, Sezione di Milano, Via Celoria 16, 20133 Milano, Italy}
\affiliation[d]{INFN, Sezione di Padova, Via Francesco Marzolo 8, 35131 Padova, Italy}
\affiliation[e]{INFN, Sezione di Torino, Via Pietro Giuria 1, 10125 Torino, Italy}
\affiliation[f]{INFN, Sezione di Ferrara, Via Saragat 1I, 44122 Ferrara, Italy}
\affiliation[g]{INFN, Sezione di Roma Tre, Via della Vasca Navale 84, 00146 Roma, Italy}

\emailAdd{ivano.sarra@lnf.infn.it}

\abstract{
The measurement of physics processes at new energy frontier experiments requires excellent spatial, time, and energy resolutions to resolve the structure of collimated high-energy jets. In a future Muon Collider, the beam-induced backgrounds (BIB) represent the main challenge in the design of the detectors and of the event reconstruction algorithms. The technology and the design of the calorimeters should be chosen to reduce the effect of the BIB, while keeping good physics performance. Several requirements can be inferred: i) high granularity to reduce the overlap of BIB particles in the same calorimeter cell; ii) excellent timing (of the order of 100 ps) to reduce the out-of-time component of the BIB; iii) longitudinal segmentation to distinguish the signal showers from the fake showers produced by the BIB;
iv) good energy resolution (less than 10\%/$\sqrt{E}$) to obtain good
physics performance, as has been already demonstrated for conceptual particle flow calorimeters.
Our proposal consists of a semi-homogeneous electromagnetic calorimeter based on lead fluoride crystals (\pbfd) read out by surface-mount UV-extended Silicon Photomultipliers (SiPMs): the Crilin calorimeter. 
In this paper, the performance of the Crilin calorimeter in the Muon Collider framework for hadron jets reconstruction has been analyzed. We report the characterisation of individual components together with the development of a small-scale prototype, consisting of 2 layers of 3 $\times$ 3 crystals each.
}

\maketitle
\flushbottom

\section{Introduction}
A Muon Collider is being proposed as a next generation facility for accelerator physics experiments. It has unique advantages both with respect to hadron colliders, permitting exact knowledge of the initial state and free from QCD background, and with respect to $e^{+}e^{-}$ colliders, because a Muon Collider can reach higher energies (due to very reduced beam bremsstrahlung) and can exploit the larger Higgs-muon Yukawa coupling. In particular, a Muon Collider can produce Higgs bosons in the s-channel with a 64 pb cross section with negligible beam radiation losses.\\
However, the Muon Collider environment is not so clean as one might expect, since the presence of the beam-induced background (BIB), produced by the decay of muons and subsequent interactions, may pose limitations on the physics performance~\cite{map}.
Although the BIB can be partially mitigated by a proper design of the machine-detector interface, for instance using two shielding tungsten nozzles in the detector region~\cite{nozzle}, it poses requirements on the detector development~\cite{bib}.\\
BIB particles at a Muon Collider have a number of characteristic features: low momentum, displaced origin and asynchronous time of arrival. The BIB flux has been simulated to be in the order of  300 $\mathrm{\gamma / cm^2}$ on the surface of the electromagnetic calorimeter (ECAL), with energy spectrum peaked around 1.8 MeV~\cite{XX1}. One of the most promising options for ECAL, proposed by the CALICE collaboration, is a sandwich of tungsten and silicon sensors~\cite{Calice} that combines a mature technology with the possibility to implement fine segmentation. The derived ECAL barrel section designed for the Compact Linear Collider (CLIC), on which the current Muon Collider detector design is based, is composed of 64 million 5$\times$5 mm$^2$ silicon sensors sampled with lead in 40 layers. Future developments should implement a precise timing measurement in these sensors ($<$100 ps) in order to make them usable at a Muon Collider. Although the high granularity is a clear advantage, the associated number of electronic read out channels is a nontrivial technological problem. Moreover, the cost for such a system exceeds the cost of other solutions. In this paper we propose a cheaper alternative as electromagnetic barrel calorimeter for the Muon Collider: Crilin, a semi-homogeneous crystal calorimeter with longitudinal information. Crilin has a modular architecture made of stackable and interchangeable sub-modules composed of matrices of lead fluoride (\pbfd) crystals, where each crystal is individually read out by two channels composed by the series of two UV-extended surface mount Silicon Photomultipliers (SiPMs) each. It can provide high response speed, good pileup capability, great light collection (hence good energy resolution) throughout the whole dynamic range, resistance to radiation and fine granularity scalable with the transverse size of the crystals. 
The longitudinal segmentation is crucial to distinguish signal showers from BIB. Adapting crystals transverse and longitudinal dimensions to maximize the signal/background ratio, our goal is to achieve similar or better performance than the silicon-tungsten ECAL proposed by CLIC. This, combined with a substantial reduction of the cost by an estimated factor of ten, could be a clear indication of the quality of our technology choice.
First studies on ECAL crystal dimensions have shown that a basic configuration with 10$\times$10$\times$40 mm$^3$ allows a good separation of BIB from signal with O(5 GeV) energy deposit per crystal. In this regard, the choice of 15 µm pixels, with the measured Light Yield (LY) of 1 p.e./MeV from an earlier existing 2-crystal prototype (Proto-0), would guarantee an excellent linearity in the response. During 2022, the construction of a bigger prototype (Proto-1), corresponding approximately to 0.7 R$_M$ (Molière radius) and 8.5 X$_0$ (radiation length), will be achieved.

\section{Crilin simulation and performance in Muon Collider environment} 

\begin{figure}[ht!]
\begin{center}
\includegraphics[width=0.49\textwidth]{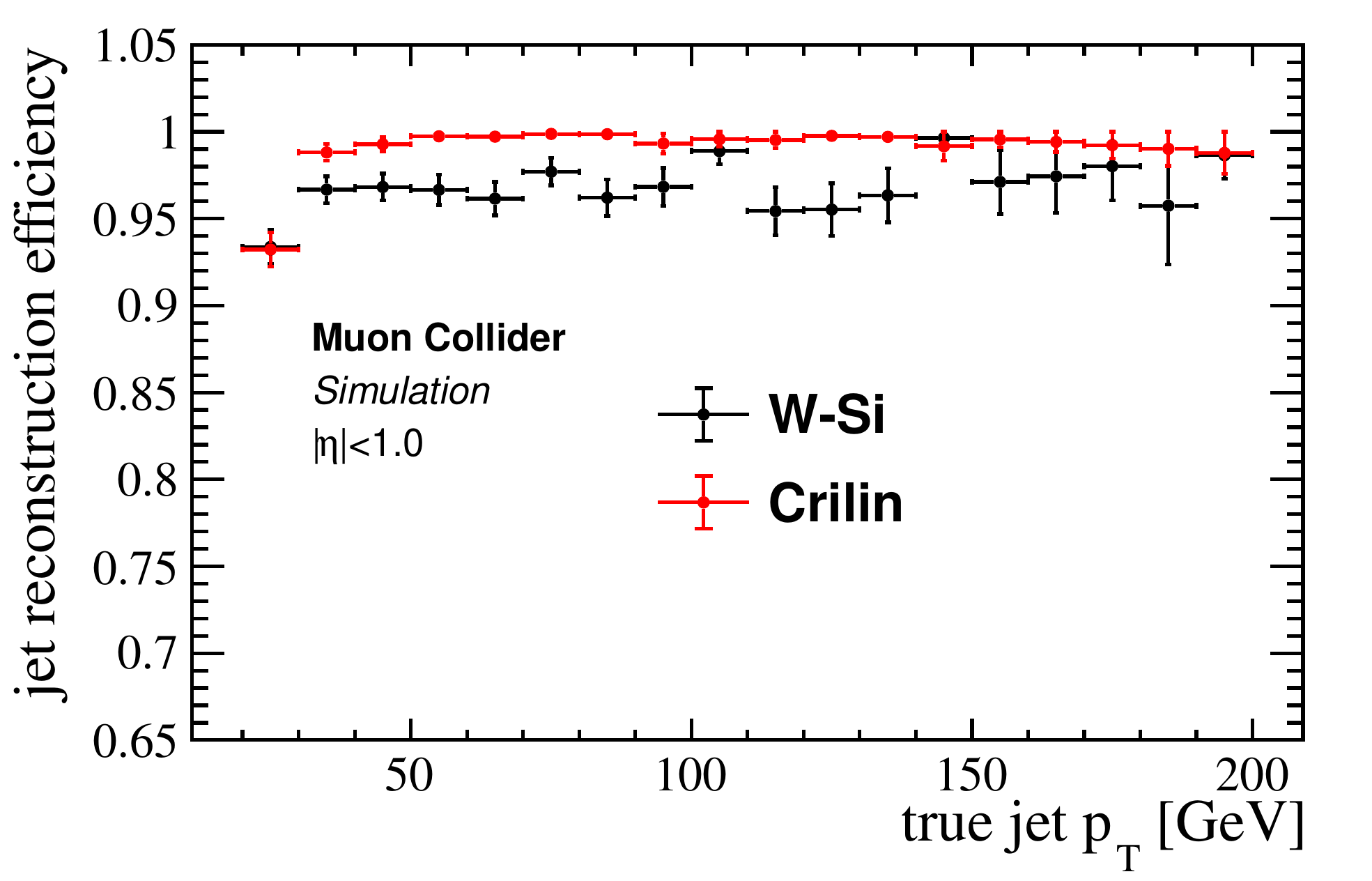}
\includegraphics[width=0.49\textwidth]{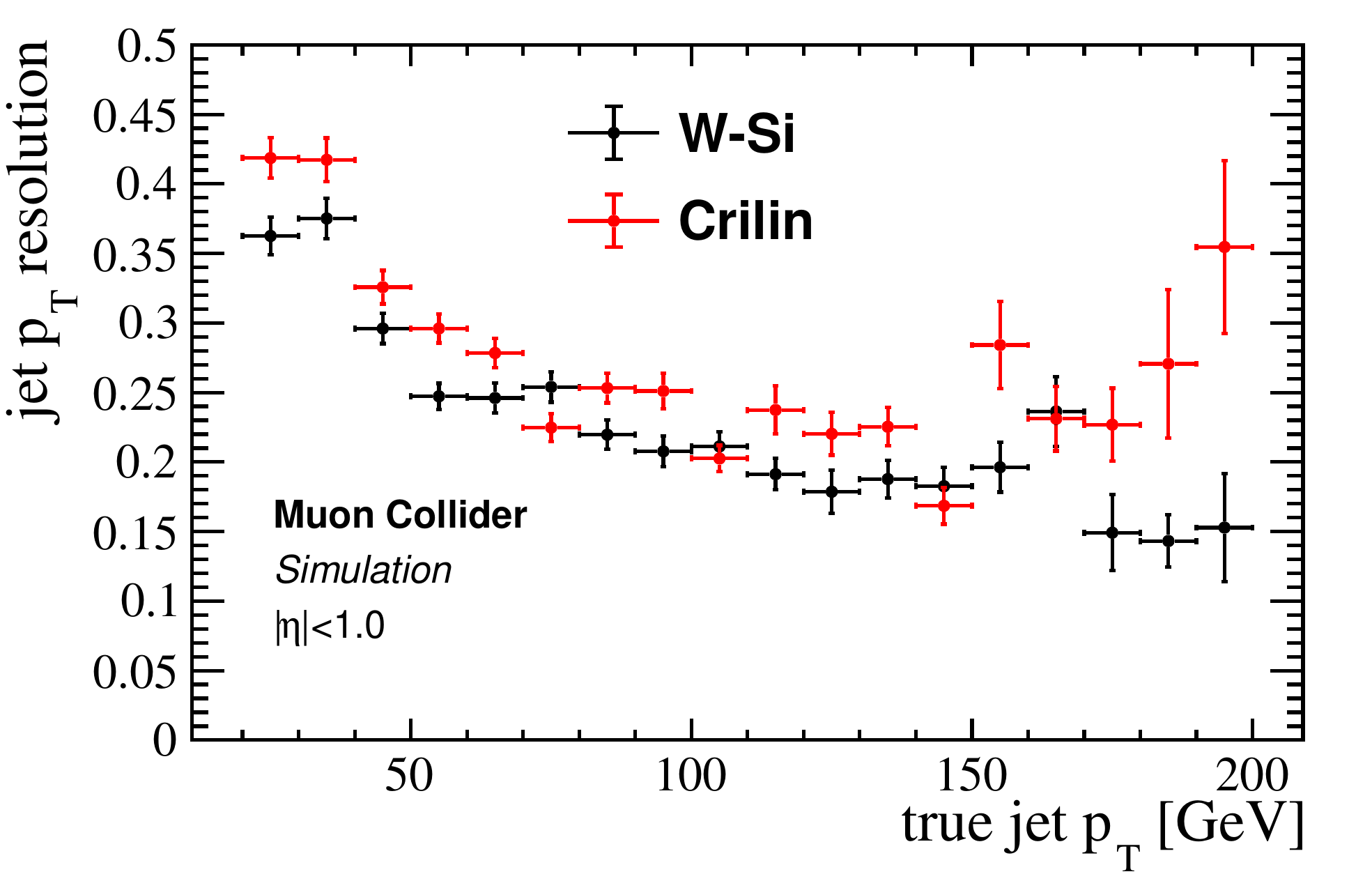}
\caption{Left: jet reconstruction efficiency as a function of the jet $p_T$, obtained by using the Crilin ECAL barrel and the W-Si ECAL barrel. Right: jet $p_T$ resolution as a function of the jet $p_T$, obtained by using the Crilin ECAL barrel and the W-Si ECAL barrel.\label{fig:jets}} 
\end{center}
\end{figure}

In order to demonstrate the viability of the Crilin technology, the simulation framework of the International Muon Collider Collaboration \cite{snowmass} has been employed.
In this study, the W-Si ECAL barrel used in this simulation framework has been substituted with the Crilin calorimeter. 
The Crilin ECAL barrel design for the Muon Collider consists of five layers of 40 mm thick each, and 10$\times$10 mm$^2$ of cell area.
The performance comparison between Crilin and W-Si calorimeters is evaluated on objects of primary interest for Muon Collider physics: hadronic jets. 
Hadronic jets are produced by the decay of the Higgs boson, and their reconstruction is fundamental for the flagship measurements at the Muon Collider: the determination of the Higgs trilinear and quadrilinear self-couplings.
The charged component of the jets, which constitute about 60\% of the jet constituents, is then precisely measured in the tracking system, leaving the energy measurement of photons (30\%) to the electromagnetic calorimeter and the measurements of neutral hadrons (10\%) to the hadron calorimeter.\\
The performance of the Muon Collider experiment with the W-Si ECAL configuration has been already analyzed in details in \cite{snowmass}, where the technique employed for the jet reconstruction is explained. In the present study the same algorithm is used for both the Crilin and W-Si detector configurations. This algorithm is not yet fully optimized, but it is useful for comparing the performance of different detector systems.
It consists of a Particle Flow algorithm that involves tracker and calorimeters~\cite{aps}. The track reconstruction is performed with the ACTS algorithm \cite{ACTS}, and includes a filter for rejecting the combinatorial tracks generated by the beam-induced background. The calorimeter clusters have been reconstructed with the PandoraPFA algorithm \cite{Pandora}, that is also used to select Particle Flow objects: tracks as charged particles, and isolated calorimeter clusters as neutral particles. Particle Flow objects then constitute the inputs to the $kt$ algorithm~\cite{KT} (with radius parameter 0.5) that performs the jet clustering. Finally a correction is applied to the jet four-momentum to take into account inefficiencies and other detector effects. It has been verified that the ECAL performance is the dominant contribution to the jet energy resolution, since tracks are precisely reconstructed, and the HCAL has a beam-induced background level much lower than the  ECAL.\\
The full simulation of a sample of $b$-jets has been used for evaluating the performance. The $b$-jet signal and the beam-induced background have been simulated for both detector configurations, with the Crilin ECAL barrel and with the W-Si ECAL barrel. In this study just the performance of jets well contained in the barrel ($|\eta|<1$, where $\eta$ is the pseudo-rapidity of the jet momentum) is considered. It has been verified that in this angular region the jet performance does not depend significantly on the jet direction. Two figures of merit are used to compare the performance of the Crilin and W-Si configurations: the jet reconstruction efficiency and jet $p_T$ resolution. The efficiency is defined as the ratio between the number of reconstructed jets and the number of truth-level jets (i.e. clustered with Monte Carlo particles). The jet $p_T$ resolution is defined as the standard deviation of the distribution of $\frac{p_T(\mathrm{jet}) - p_T(\mathrm{MC-jet})}{p_T(\mathrm{MC-jet})}$, where $p_T(\mathrm{jet})$ is the transverse momentum of the reconstructed jet, and $p_T(\mathrm{MC-jet})$ is the transverse momentum of the associated truth-level jet. It has also been verified that the fake jet rate is comparable in the two calorimeter configurations, in order to perform a proper comparison. The jet reconstruction efficiency and jet $p_T$ resolutions are presented in Figure \ref{fig:jets}. It can be seen that the performance is similar in the two cases, where the W-Si or the Crilin ECAL barrel are used.

Minor differences are observed at true $p_T$ greater than 170 GeV. The Crilin resolution degradation is due to the lower number of radiation lengths ($\sim19$ X$_0$) respect to the one of the W-Si ($\sim22$ X$_0$). 

\section{Radiation environment and radiation damage}\label{Sec:sec1}
The characterization of beam-induced background is of paramount importance to quantify the requirements on the detector at a Muon Collider. The BIB at $\sqrt{s} = 1.5$ TeV has been simulated by means of FLUKA~\cite{fluka}. The results are analyzed and benchmarked against the ones obtained with the code MARS15 by the Muon Accelerator Program (MAP) collaboration in Ref.~\cite{Collamati:2021sbv}.
The FLUKA simulation to obtain the dose maps, not including the contribution from the muon collisions, employs a simplified detector geometry: all the silicon layers composing the inner tracker are included with exact dimensions. The calorimeters, magnetic coils, and the return yoke are approximated with cylindrical elements with densities and material composition based on the averages from the full geometry. The magnetic fields are assumed to be uniform.

\begin{figure}[ht!]
    \begin{center}
        \centering
        \includegraphics[width=.7\textwidth]{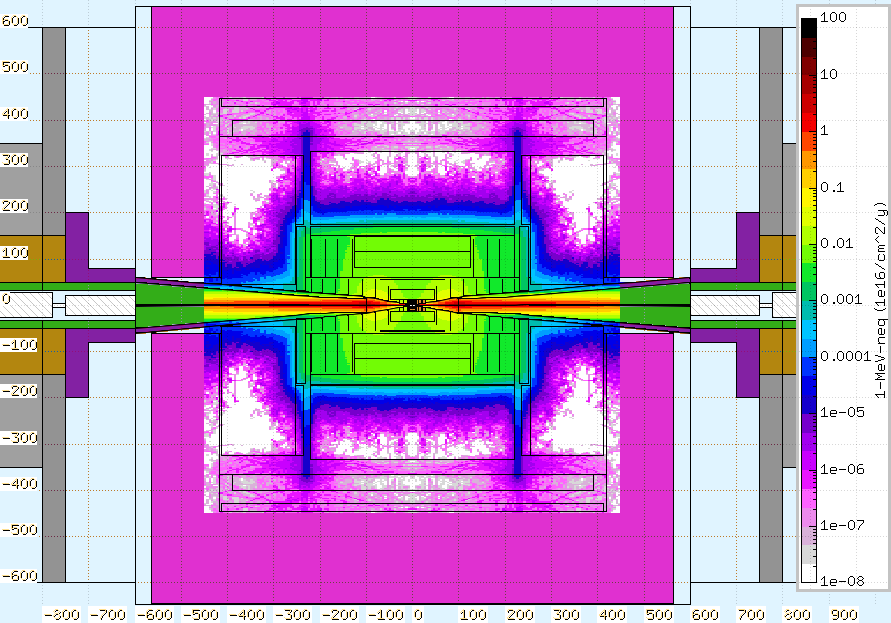}
    \end{center}
    \caption{Map of the 1-MeV-neq fluence in the detector region for a Muon Collider operating at $\sqrt{s} = 1.5$~TeV with the parameters described in Ref.~\cite{Collamati:2021sbv}, shown as a function of the position along the beam axis and the radius. The map is normalised to one year of operation (200 days/year) for a 2.5 km circumference ring with 5 Hz injection frequency.}
    \label{fig:fluence}
\end{figure}

\begin{figure}[ht!]
    \begin{center}
        \centering
        \includegraphics[width=.7\textwidth]{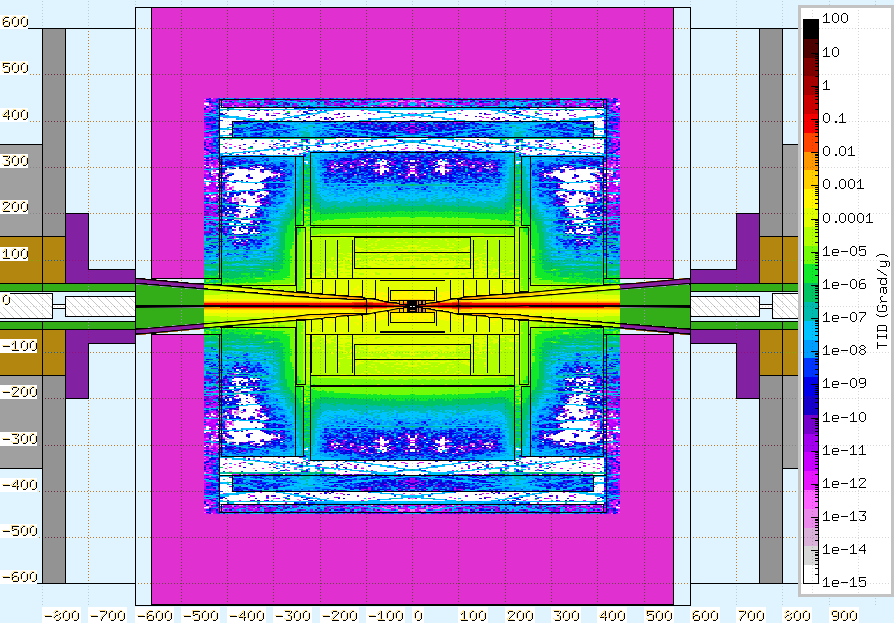}
    \end{center}
    \caption{Map of the TID in the detector region for a Muon Collider operating at $\sqrt{s} = 1.5$~TeV with the parameters described in Ref.~\cite{Collamati:2021sbv}, shown as a function of the position along the beam axis and the radius. The map is normalised to one year of operation (200 days/year) for a 2.5 km circumference ring with 5 Hz injection frequency.}
    \label{fig:tid}
\end{figure}

Figures~\ref{fig:fluence} and~\ref{fig:tid} show respectively the expected 1 MeV neutron equivalent fluence (1-MeV-neq) and the total ionizing dose (TID) in the detector region, shown as a function of position along the beam axis z and the radial distance $r$ from the beam axis. The normalization for the dose maps is computed considering a 2.5 km circumference ring and an injection frequency of 5 Hz. 
Assuming 200 days of operation during a year, the 1-MeV-neq fluence is expected to be $\sim 10^{14-15}$ cm$^{-2}$y$^{-1}$ in the region of the tracking detector and of $\sim 10^{14}$ cm$^{-2}$y$^{-1}$ in the electromagnetic calorimeter, with a steeply decreasing radial dependence beyond it. The total ionizing dose is $\sim10^{-3}$ Grad/y on the tracking system and $\sim10^{-4}$ Grad/y on the electromagnetic calorimeter.

Preliminary FLUKA simulations at $\sqrt{s} = 3$ TeV and $\sqrt{s} = 10$ TeV present a BIB of the same intensity level as for $\sqrt{s} = 1.5$ TeV. An accurate optimization at those energies can possibly lead to a further improvement of the results.\\

\section {Crystal Characterization}\label{Sec:sec2}

\begin{figure}[ht!]
    \centering
    \includegraphics[width=0.6\textwidth]{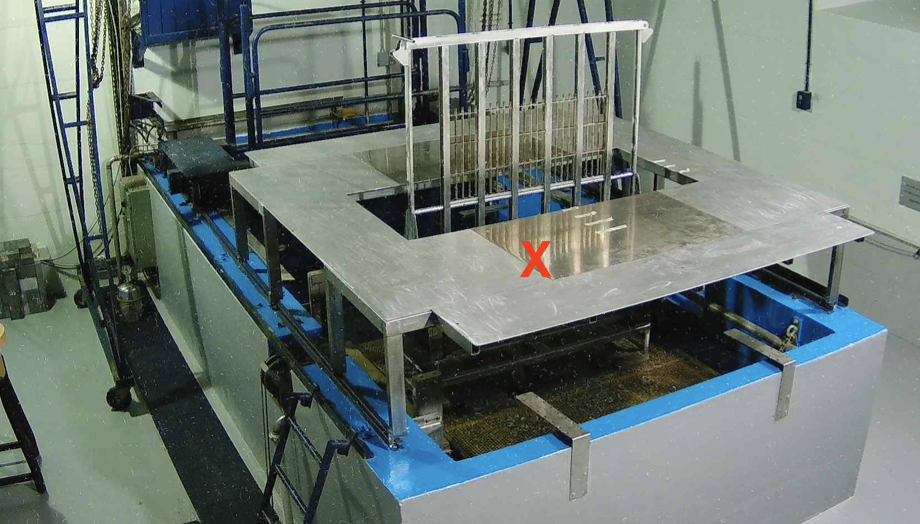}
    \caption{Setup during the irradiation phase inside the Calliope facility; the red cross represents the position of the crystals during the irradiation.}
    \label{fig:Gamma_Irr_setup}
\end{figure}

The characterization campaign of \pbfd crystals was started in February 2021: the individual and combined effect of TID and neutrons was evaluated by measuring the resulting deterioration in transmittance for two crystals sized $5\times5\times40 $ mm$^3$, manufactured by SICCAS~\cite{SIC} using a melt growth process, thus resulting in a cubic form ($\beta$-PbF$_2$). 
The first crystal was tested without any kind of wrapping (in the following referred to as the ``naked'' one); the other was wrapped with a 100 $\mu$m thick Mylar foil, which reflects the crystals wrapping choice for Proto-1. A detailed description of the measurements performed can be found in \cite{articolo_irraggiamenti}. 
After reference transmittance measurements, the irradiation phase was carried out at Calliope~\cite{Calliope}, a pool-type gamma irradiation facility equipped with a $^{60}$Co radio-isotopic source array housed in a large volume $(7.0\times 6.0 \times 3.9$ m$^3$) shielded cell. The Calliope source rack is composed of 25 $^{60}$Co source rods (with  $41 \times 90$ cm$^2$ active area) arranged in a planar geometry, producing photons with $E_{\gamma} = 1.25$ MeV; the activity of the plant was $1.97\times 10^{15}$ Bq. As shown in \figurename~\ref{fig:Gamma_Irr_setup}, the crystals were positioned 70 cm away from the source, with their longitudinal axes perpendicular to it, yielding a 100 krad/h dose rate.\\

\begin{figure}[ht!]
    \centering
     \begin{tabular}{cc} 
    \includegraphics[width=0.49\textwidth]{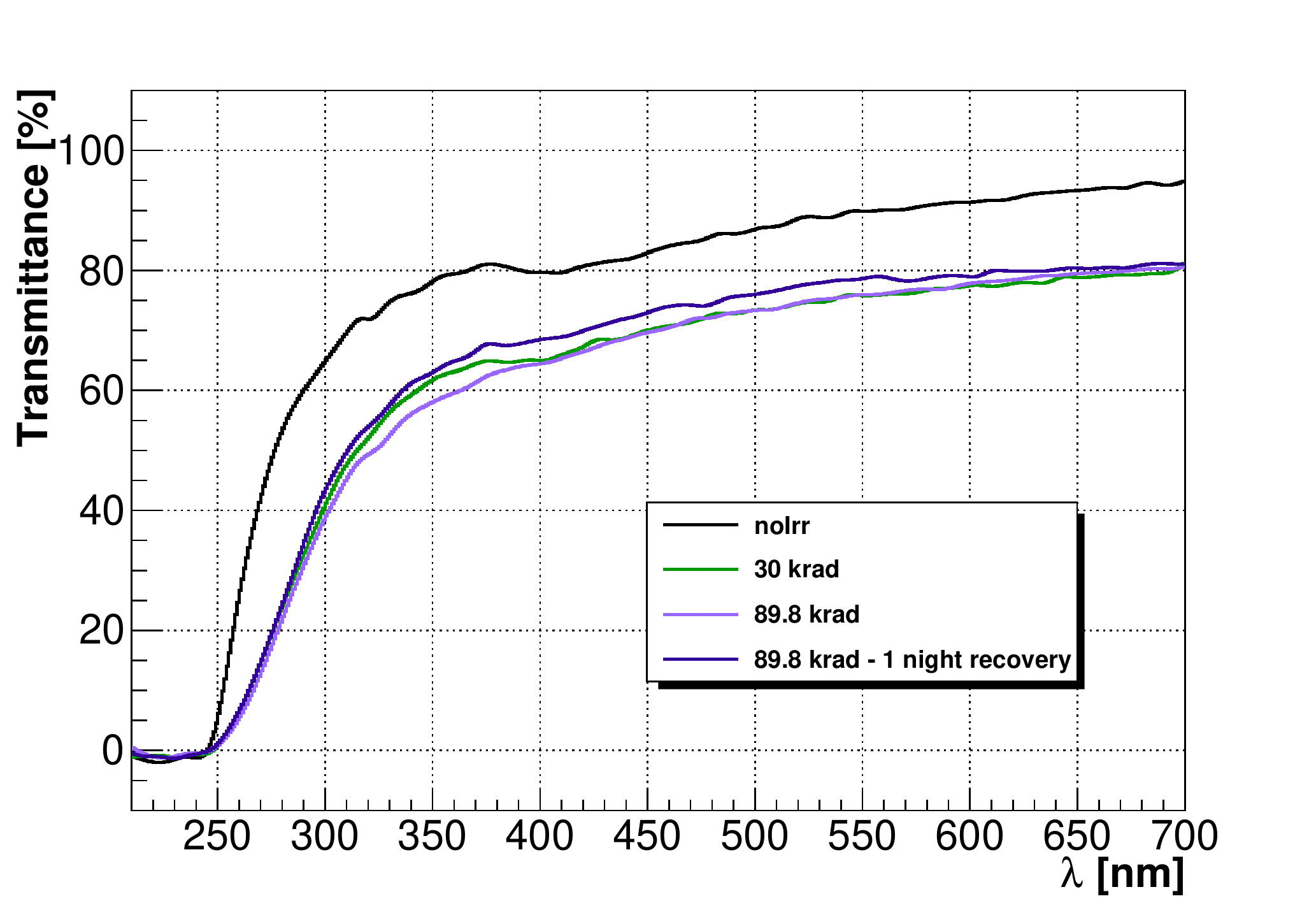}
    \includegraphics[width=0.49 \textwidth]{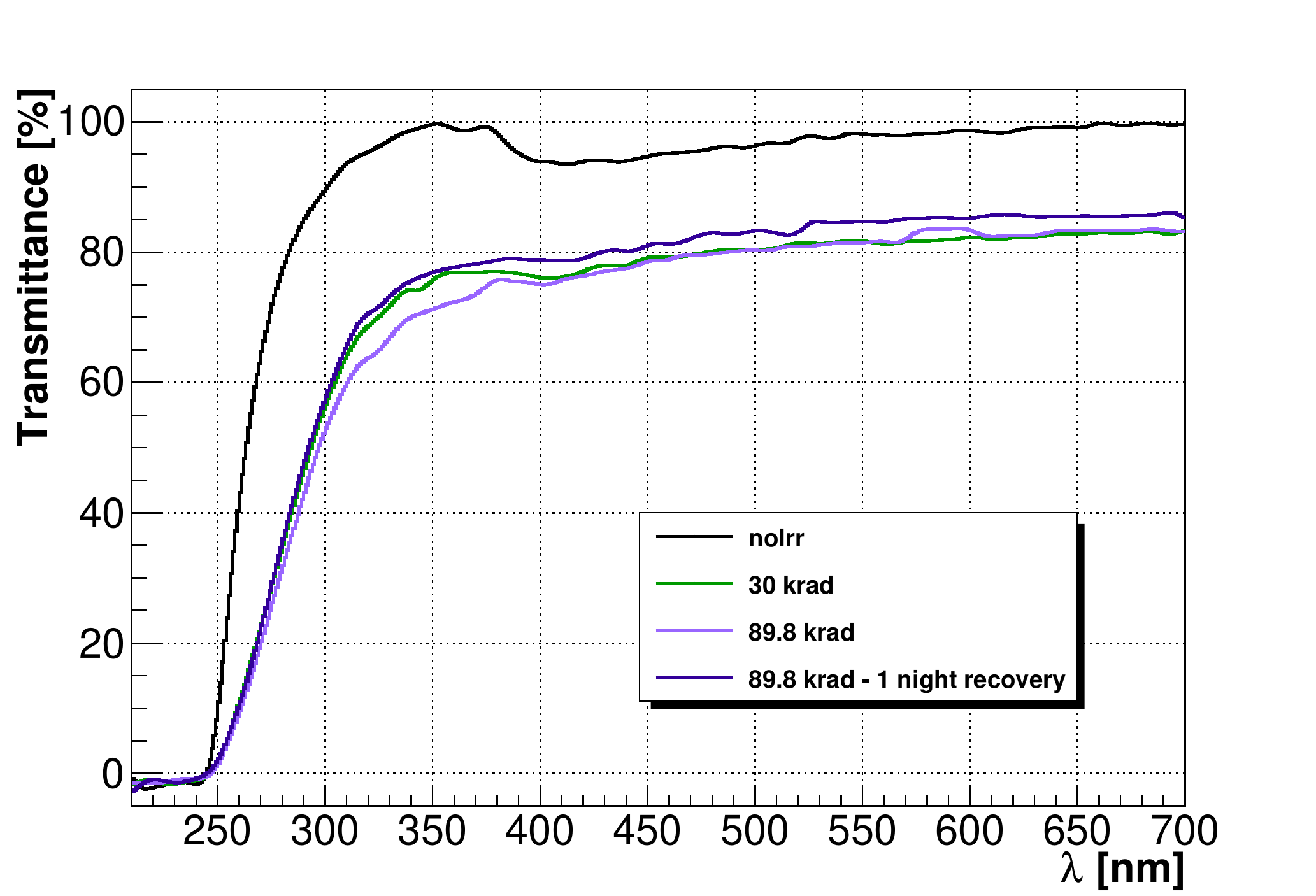}\\
     \includegraphics[width=0.49 \textwidth]{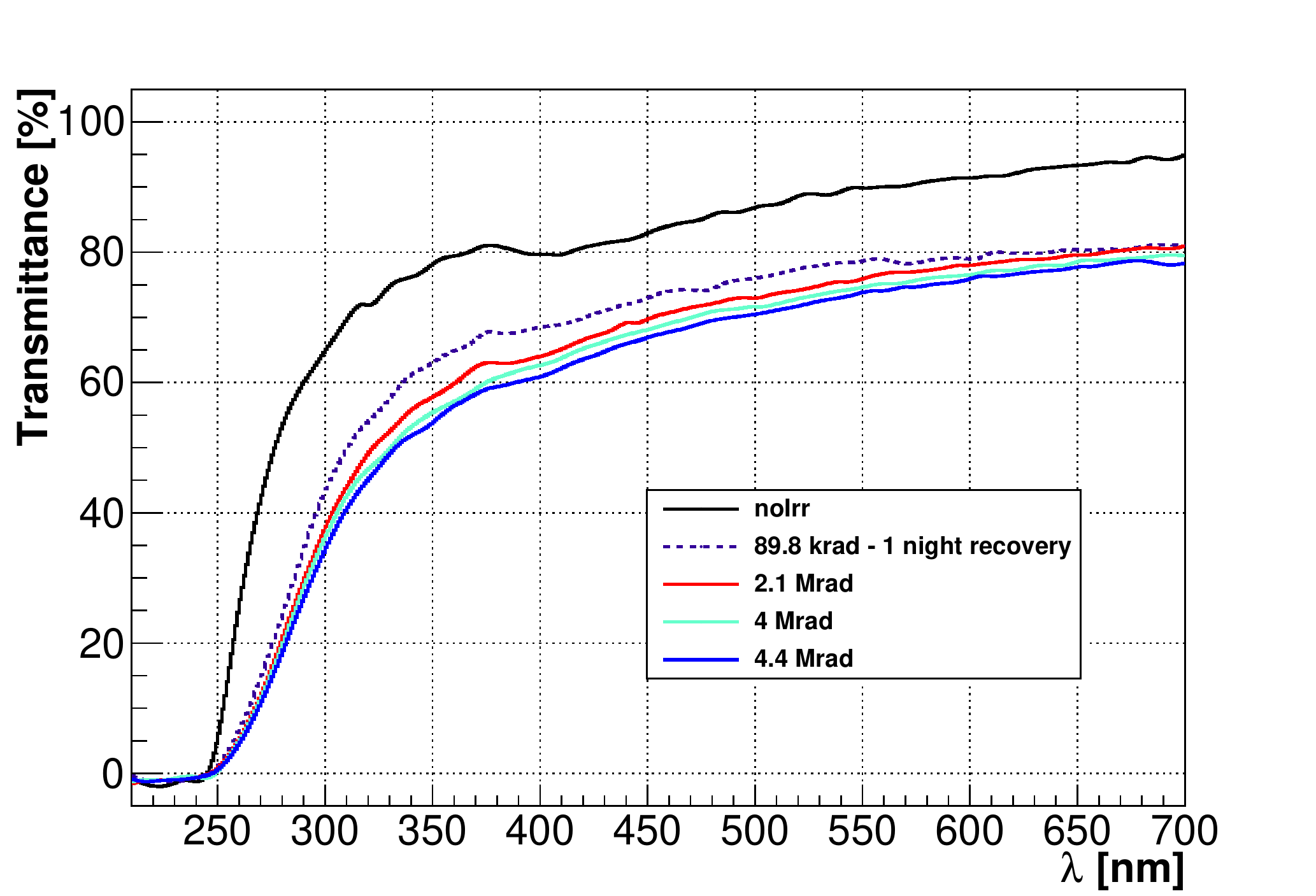} \includegraphics[width=0.49\textwidth]{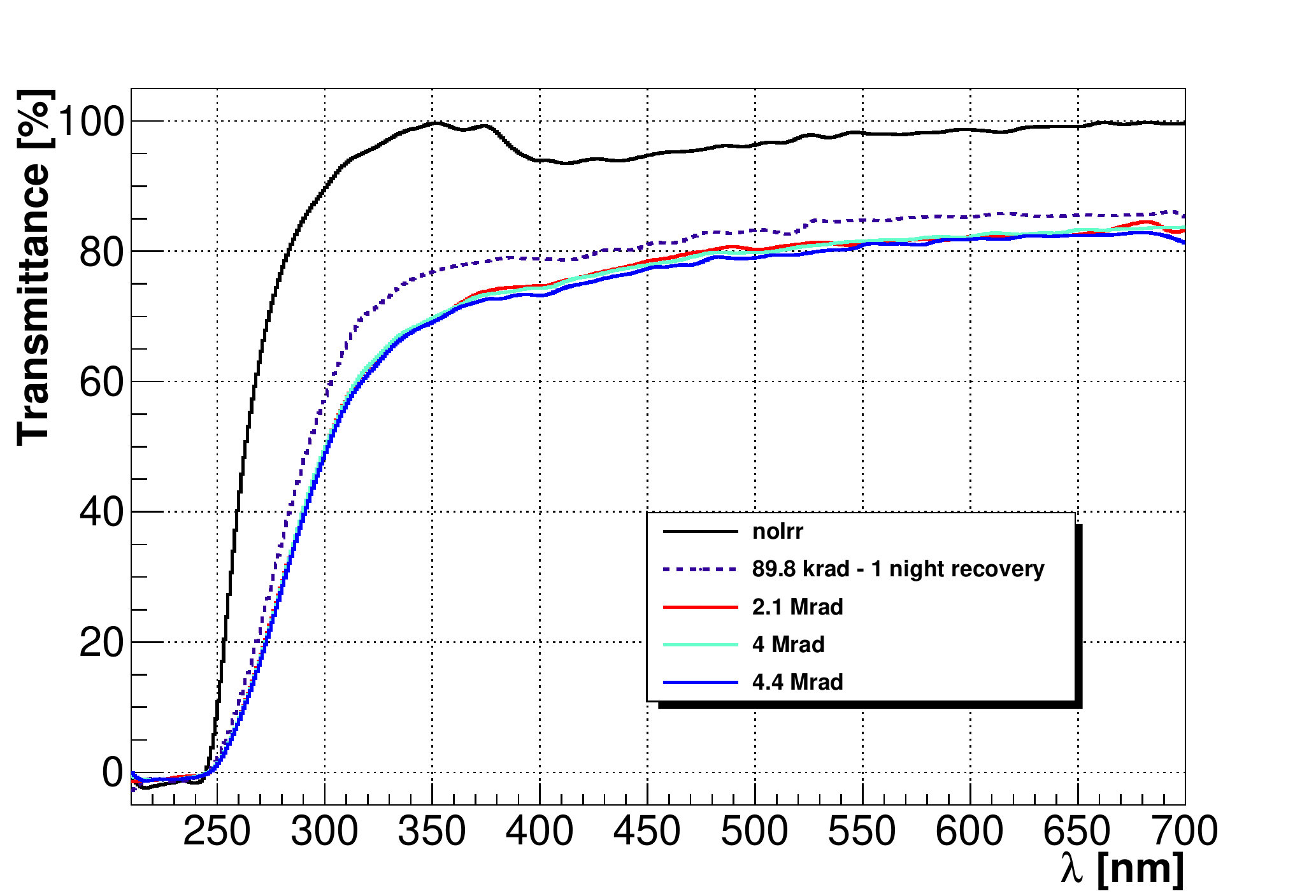}
    \end{tabular}
    \caption{Transmission spectra obtained in the different irradiation steps  for the naked crystal (top and bottom left) and the crystal with Mylar wrapping (top and bottom right).}
    \label{fig:trasm_1_to_5}
\end{figure}

The transmittance measurements were performed at different irradiation steps during three days. 
In Table~\ref{tab:Irrstep} the absorbed doses, expressed in krad (air),  at each irradiation step are reported.
\figurename~\ref{fig:trasm_1_to_5} shows the longitudinal transmittance spectra obtained at different irradiation steps for the two crystals.
After a TID of approximately 80 krad, further exposure does not lead to significant reductions in transmittance, suggesting a saturation effect associated with the damage mechanism, in accordance with what is reported in~\cite{zhu1996study}.\\
\begin{table}[ht!]
    \centering
    \begin{tabular}{|c|c|} 
    \hline
    Irradiation Step & Dose in air [krad] \\ \hline
   I & 30.2 \\
   II & 89.88\\
   III & 2082 \\
   IV & 4031.8 \\
   V & 4435.5\\ \hline
   \end{tabular}
    \caption{Irradiation steps and corresponding total dose absorbed by the crystals.}
    \label{tab:Irrstep}
\end{table}

\begin{figure}[ht!]
    \centering
    \includegraphics[width=0.6\textwidth]{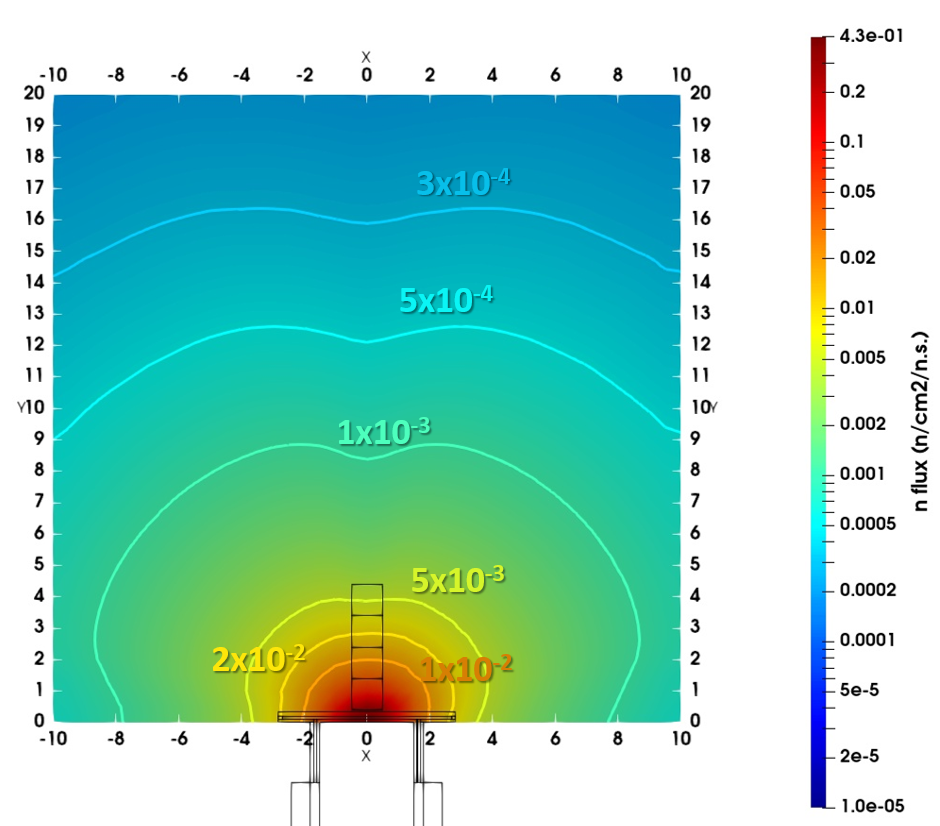}
    \caption{Simulation of the neutron flux per produced neutrons per second: crystals were placed 1 cm away from the neutron gun.} 
    \label{fig:Simu_Neutron}
\end{figure}

The maximum degradation observed is at the level of $\sim40\%$, within the range of degradation observed in~\cite{ren2001optical} and~\cite{achenbach1998radiation}.
 After the irradiation phase, crystals were kept in a dark box to allow their natural recovery, according to the procedure described in~\cite{kozma2002radiation}.
 Moreover, to further improve the recovery process, a light-bleaching run with 400 nm blue light was performed according to the procedures and methods described in~\cite{achenbach1998radiation}.
 A few percent recovery was observed in the transmittance spectrum after 16 hours, indicating a small accelerating effect on the recovery process with respect to the natural annealing case.
Subsequently, the crystals were irradiated at the Frascati Neutron Generator (FNG) facility of ENEA Frascati \cite{FNG}.\\
Neutron generation at FNG is based on the T(d,n)$\alpha$ fusion reaction, producing 14 MeV neutrons with a flux up to $10^{12}$ neutrons/s in steady state or pulsed mode. The front face of the crystals were positioned 1 cm away from the source. \figurename~\ref{fig:Simu_Neutron} reports the neutron flux map in the FNG facility. To  evaluate the expected neutron fluence, the irradiation process was implemented on the McStas simulation package \cite{mcstas}, using 1 cm bins along the crystal axes.

In Table~\ref{tab:NeutronIrrstep} the instantaneous flux per produced neutrons per second\footnote{Corresponding to a total of $1 \times 10^{12} $ produced neutrons per second.} and the fluence, integrated in 1 hour and 30 minutes of irradiation, are reported for the 4 cm of the crystal. In the first cm of the crystals the total fluence was $10^{13}$ n/cm$^2$, while in the fourth cm it was $4.98 \times 10^{11}$: a reduction of a factor $\sim$20 in the fluence is observed at the two edges of the crystal.
\begin{table}[ht!]
    \centering
    \begin{tabular}{|c|c|c|} 
    \hline
    Crystal Sector [cm] & Flux [n/(cm$^2$)/n.s.] & Fluence [n/cm$^2$]\\ \hline
   1 & $9.35 \times 10^{-2}$ & $1.01 \times 10^{13}$\\
   2 & $2.22 \times 10^{-2}$ & $2.40 \times 10^{12}$\\
   3 & $9.03 \times 10^{-3}$ & $9.76 \times 10^{11}$\\
   4 &  $4.61 \times 10^{-3}$ & $4.98 \times 10^{11}$\\ \hline
   \end{tabular}
    \caption{Average of the flux and total fluence per each crystal sector: a sector correspond to 1 cm length of crystal along the longitudinal axis.}
    \label{tab:NeutronIrrstep}
\end{table}

\begin{figure}[ht!]
    \centering
     \begin{tabular}{cc} 
    \includegraphics[width=0.49 \textwidth]{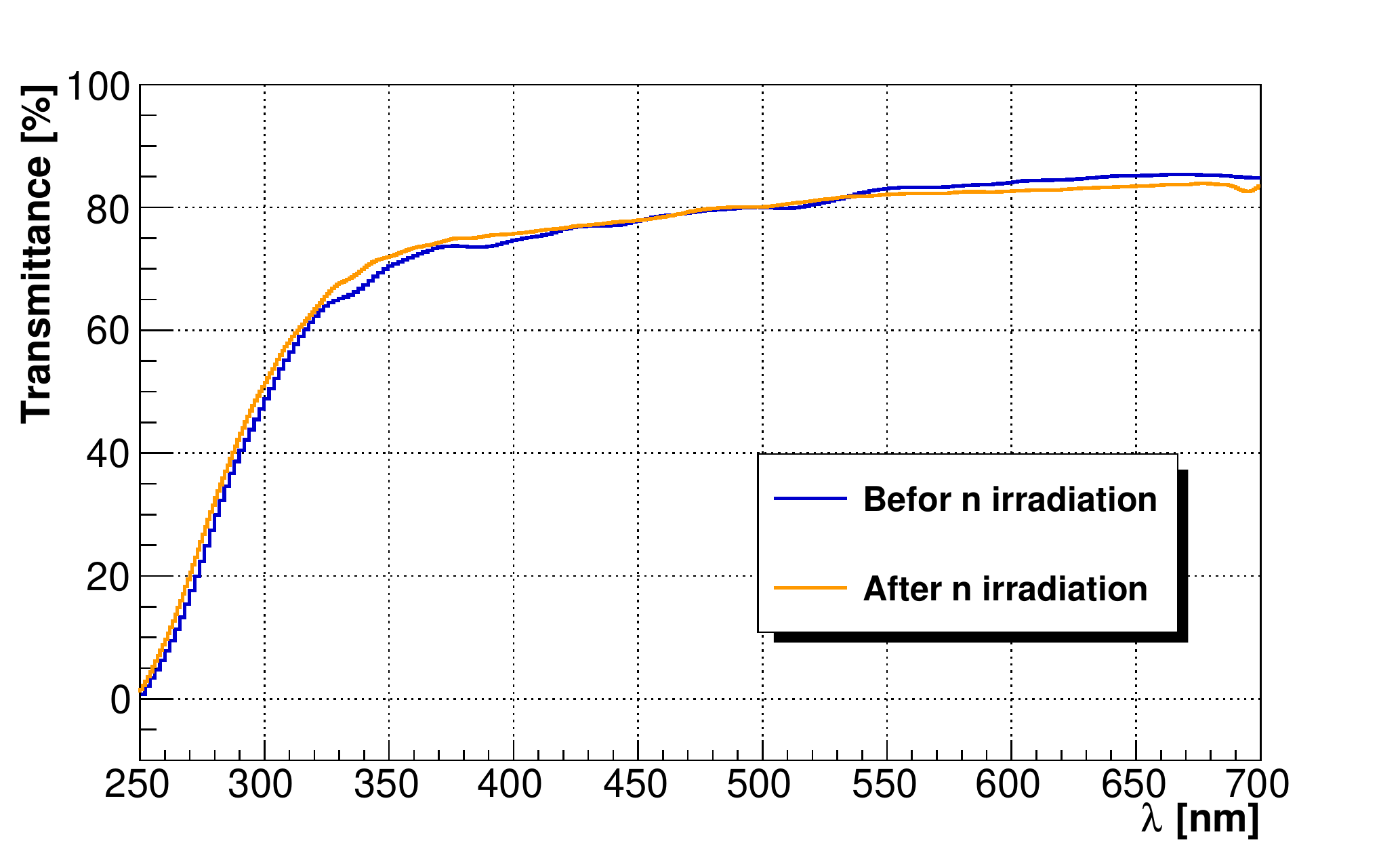} & 
    \includegraphics[width=0.49 \textwidth]{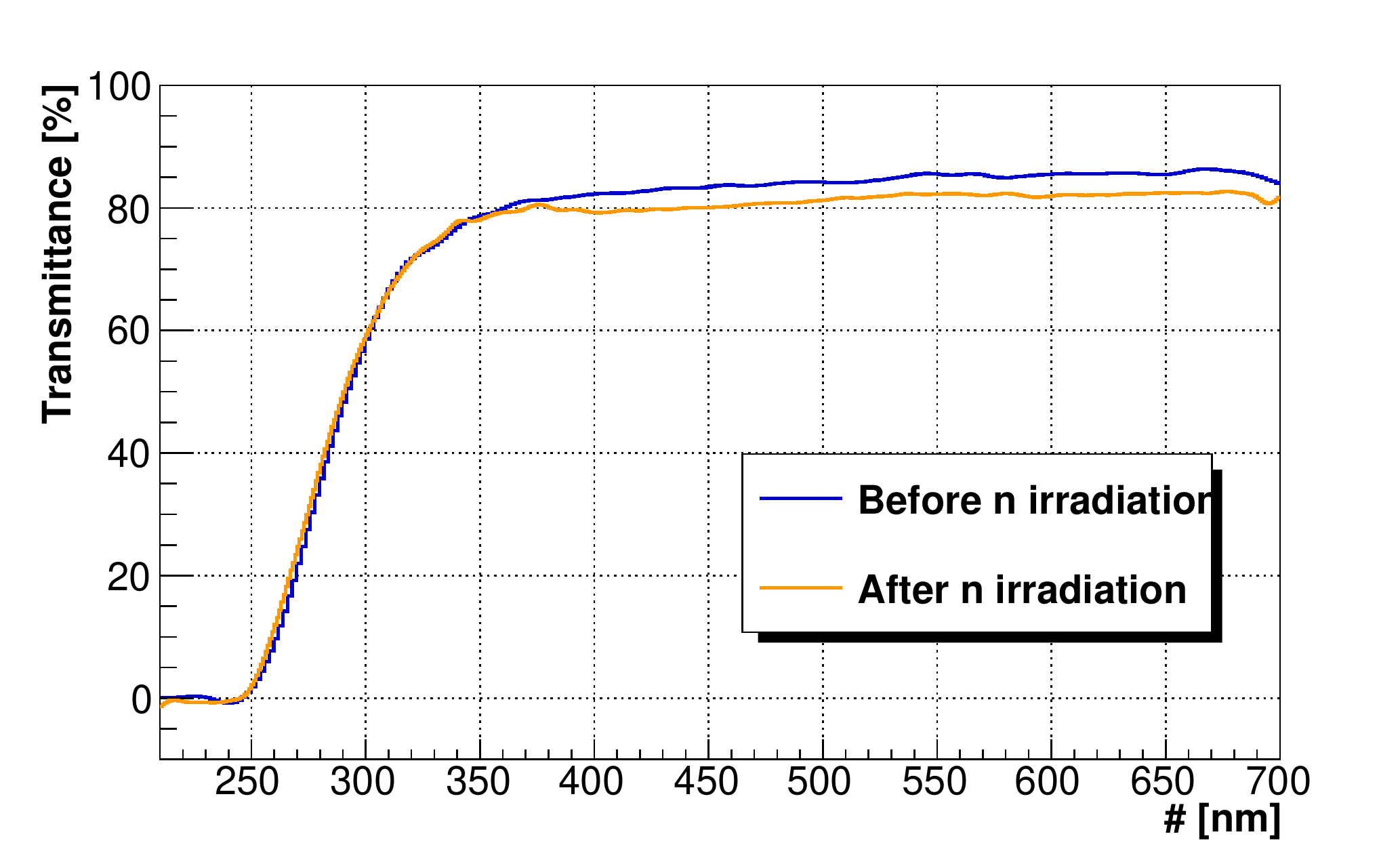} 
    \end{tabular}
    \caption{Transmission spectra obtained after the  irradiation at FNG with 14 MeV neutrons for a total fluence of $10^{13}$\;n/cm$^2$ (orange line) compared with the results  after the 16 hours optical bleaching (blue line) for the naked crystal (left) and for the crystal with Mylar wrapping (right).}
    \label{fig:DopoNeutroni}
\end{figure}

Because of technical time related to logistics and shipment of the crystals, transmittance measurements could only be performed 14 days after the irradiation and showed no deterioration in the transmittance spectrum as reported in \figurename~\ref{fig:DopoNeutroni}.

     \label{fig:fig2}
% \end{figure}

\section{Crilin Prototype}\label{Sec:sec3}
\begin{figure}[ht!]
    \begin{center}
        \centering
        \includegraphics[width=.8\textwidth]{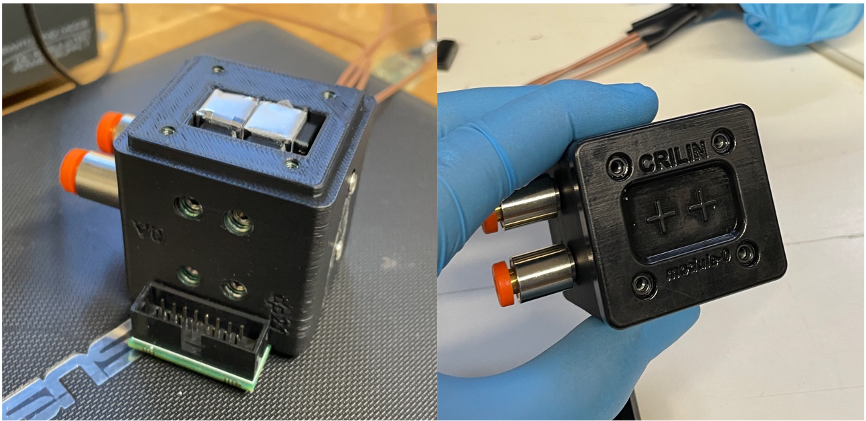}
    \end{center}
    \caption{Proto-0: two crystals prototype built in 2021.}
    \label{fig:proto0.png}
\end{figure}

The novel idea behind the Crilin ECAL is to use multiple layers of \pbfd crystals and thin surface-mount device (SMD) SiPMs stacked on top to provide longitudinal information. One layer with 2 crystals (Proto-0) has been built in 2021, as shown in Figure~\ref{fig:proto0.png}, and tested at the Beam Test Facility of Laboratori Nazionali of Frascati with 500 MeV electrons in July 2021 and at the H2 test facility of CERN with 120 GeV electrons in August 2021~\cite{MATT}.

Promising results in terms of time resolution have been achieved: less than 100 ps for deposited energies greater than 1 GeV, and about 1 p.e./MeV of light yield. With Proto-0 we tested the timing performances and set the requirements for electronics and cooling. In order to validate the design choices, the proposal is to build a larger prototype, called Proto-1. The design will be optimized with the simulation studies  starting from  dimensions of 0.7 R$_M$ and 8.5 X$_0$ ($\sim$0.3 $\lambda$). This size comes from a compromise of an acceptable containment of ~100 GeV electrons and cost constraints. Results will be extrapolated to the optimum length of the Muon Collider calorimeter of the order of 20 X$_0$. The proposal is to build Proto-1 with two layers of 3x3 \pbfd crystals, each read out with UV-extended SiPMs (Hamamatsu S14160-3015PS SMD sensors~\cite{SiPMdatasheet}), as already done in Proto-0. These new SiPMs were already tested with an ultra-fast blue laser (400 nm, 100 ps) and new electronics front-end (FEE) that showed a dynamic range from 0 to 2V, a rise time of $\sim$2 ns with full signal in $\sim$70 ns and a $\sigma_t$ less than 50 ps even at charge as low as 100pC ($\sim$250 Np.e., see Section~\ref{Subsec:Subsec5}).\\
Proto-1 operational temperature will be 0/-10 \textdegree C and the performance will be validated in a dedicated test beam. 
Specifically, our goals are: 1) perform a complete operational test of the prototype, including operation with cooling; 2) obtain data for a complete analysis of digitized signals from the detector for electrons and minimum-ionising particles; 3) test the cluster reconstruction capability and measure the time resolution; 4) measure longitudinal and transverse shower profile and compare with results obtained in simulation.

\subsection{Electronics}\label{Subsec:Subsec3}
\begin{figure}[ht!]
    \begin{center}
        \centering
        \includegraphics[width=.8\textwidth]{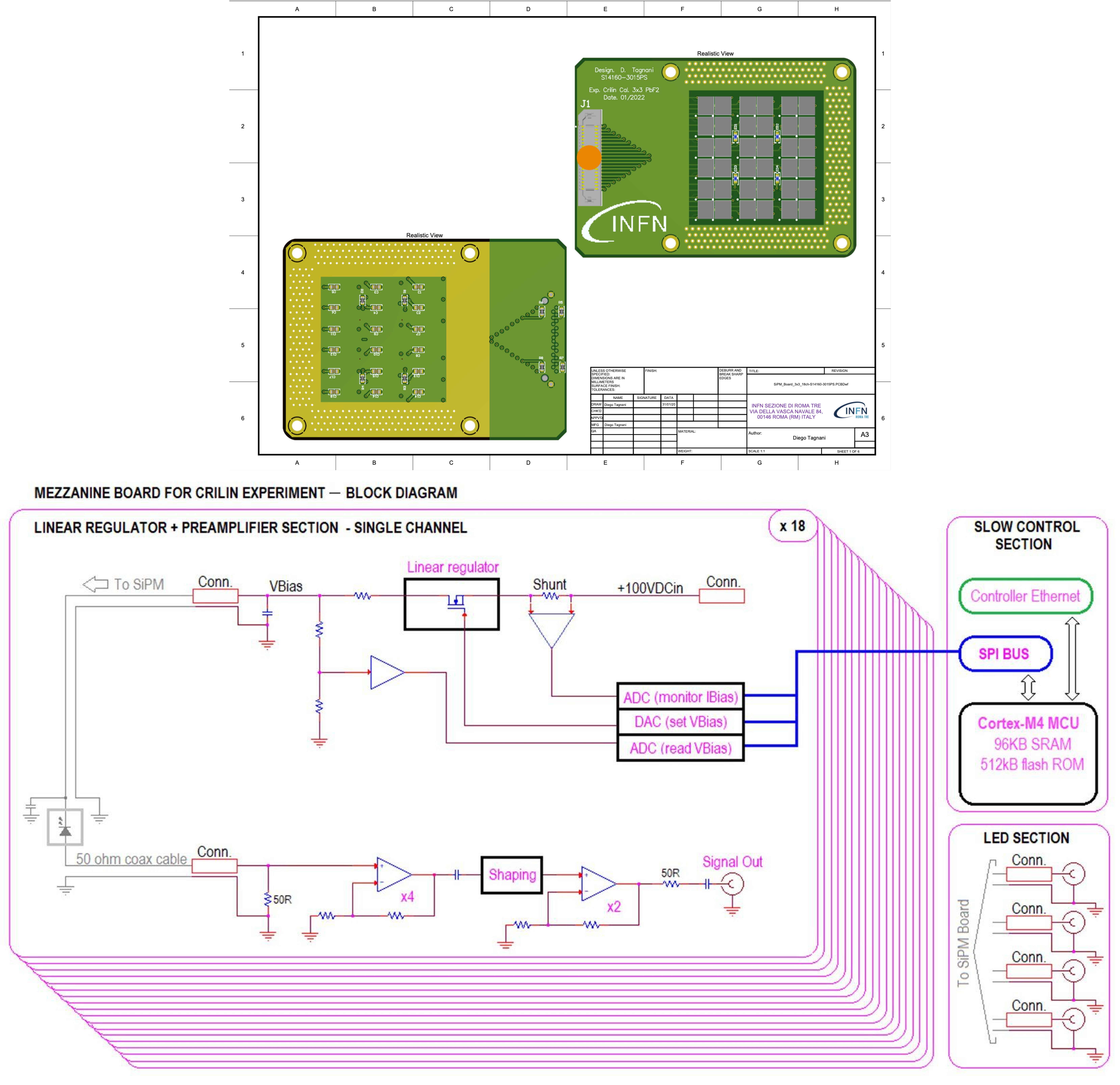}
    \end{center}
    \caption{SiPM board CAD (top) and Mezzanine Boards CAD rendering (bottom).}
    \label{fig:elec1.png}
\end{figure}

\begin{figure}[ht!]
    \begin{center}
        \centering
        \includegraphics[width=.8\textwidth]{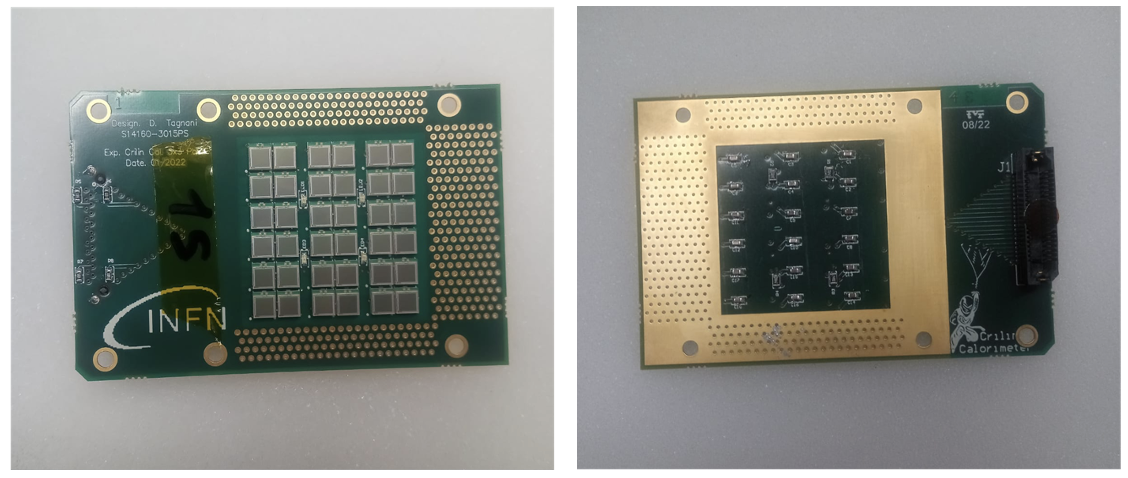}
    \end{center}
    \caption{Produced SiPM board for Proto-1.}
    \label{fig:elec2.png}
\end{figure}

Crilin’s FE electronics will be composed of two subsystems: the SiPM board (Figure~\ref{fig:elec1.png}, top) and the Mezzanine Board (Figure~\ref{fig:elec1.png}, bottom). The SiPM board  houses a layer of 36 photo-sensors, so that each crystal in the matrix is equipped with two separate and independent readout channels, the latter being composed of a series connection of two Hamamatsu S14160-3015PS SMD sensors. The 15-um SiPM pixel size, along with the series connection of two photo-sensors, were selected for high-speed response, short pulse width and to better cope with the expected total non-ionising dose (TNID) without showing an unmanageable increase in bias current during operation. The series connection of two photo-sensors further contributes to achieving a high-speed response, by effectively halving the photo-sensor equivalent capacitance. A group of four 0603 SMD blue LEDs are installed on the SiPM board, nested between the photo-sensor packages. They will be driven by means of an external nanosecond pulse generator, to allow in-situ calibration, diagnostics, and monitoring for all photo-sensors in the matrix. In Figure~\ref{fig:elec2.png} the produced SiPM boards are shown.\\ 

\begin{figure}[ht!]
    \centering
    \begin{tabular}{cc} 
    \includegraphics[width=0.48 \textwidth]{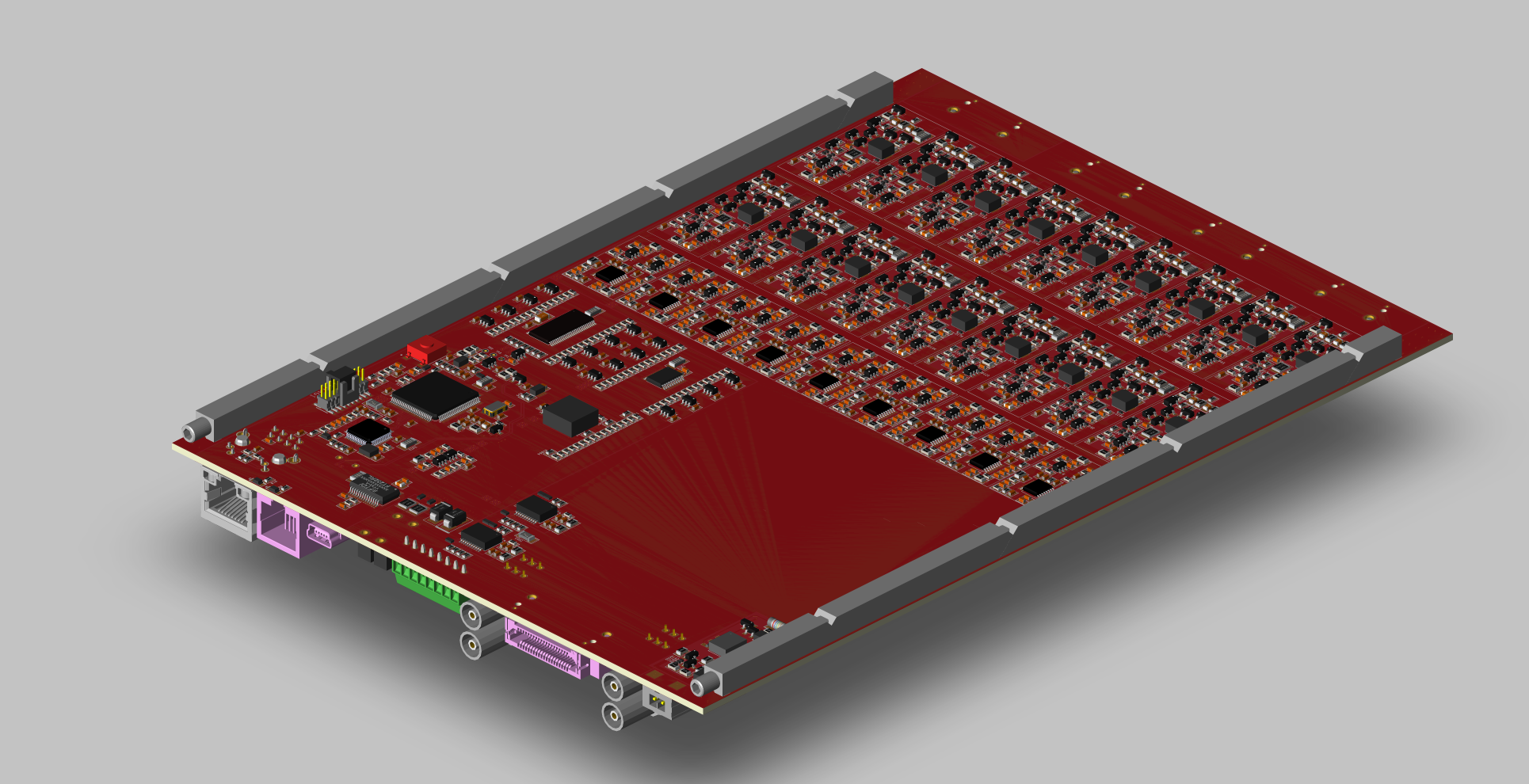} & 
    \includegraphics[width=0.48 \textwidth]{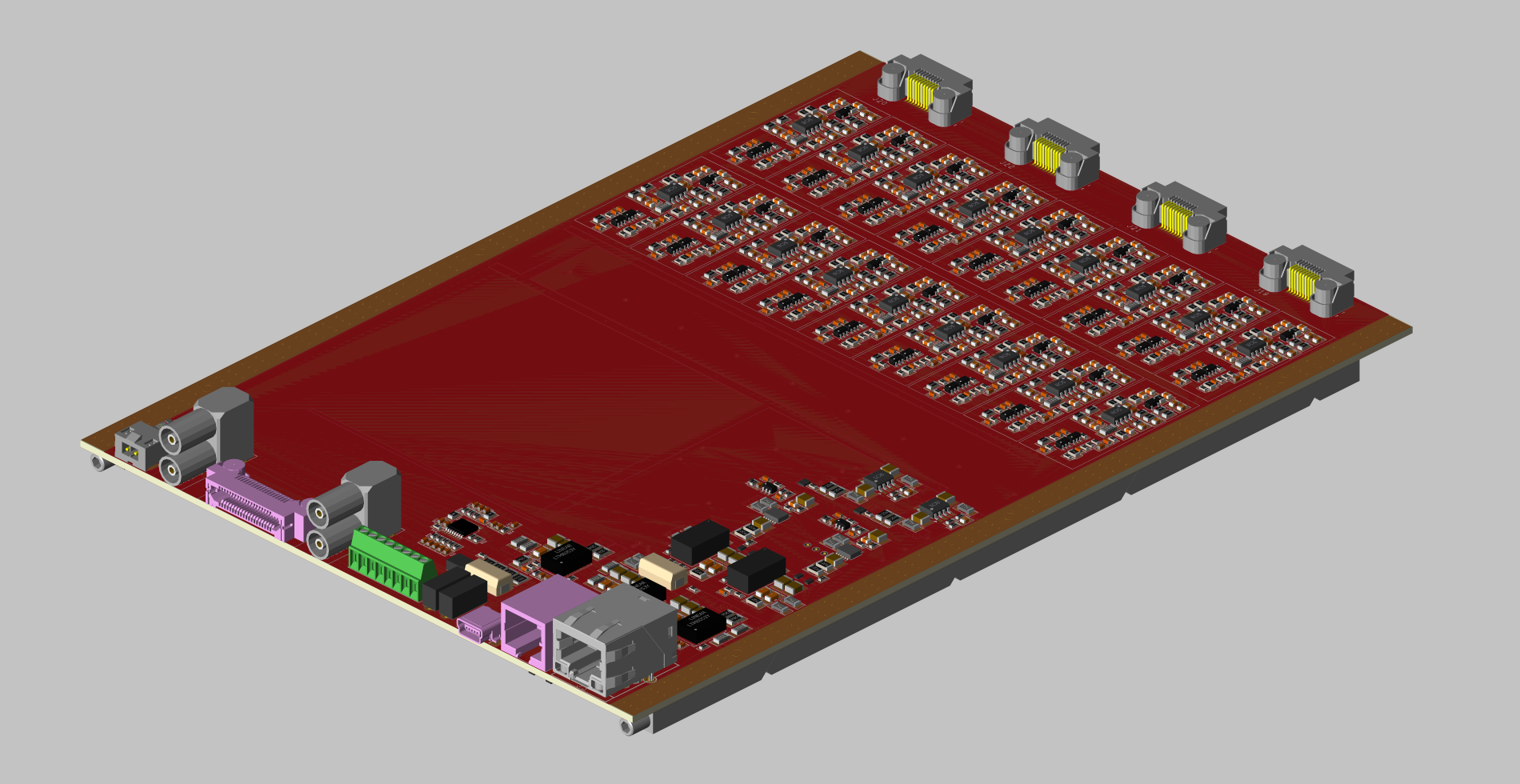} 
    \end{tabular}
    \caption{3D model of the Crilin mezzanine board: top (left picture) and bottom (right picture) layers.}
    \label{fig:3Delettronica}
\end{figure}

All bias voltages and SiPM signals for each readout channel are transported between the SiPM board and the Mezzanine Board by means of individual 50-ohm micro-coaxial transmission lines (Samtec ERCD-040-40.00-TEU-TED-1-D~\cite{SAMTEC}). Decoupling capacitors for each channel, along with a PT1000 temperature sensor, are also installed onboard.\\
The Mezzanine Board provides signal amplification and shaping, along with all slow control functions, for a total of 18 readout channels. Signals transmitted through the micro-coax lines, after proper termination, are amplified first by a high-speed non-inverting stage with gain 4. The first stage drives a pole-zero cancellation circuit, followed by a second non-inverting stage, which drives the digitisation section with a dynamic range of 2 V and an overall gain of 8. An example of a digitised SiPM waveform is reported in Figure~\ref{fig:waveformSiPMs}.
SiPM biases are generated on-board by means of 18 high-voltage linear regulators with a 0-100 V dynamic range and a maximum current sourcing capability of 5 mA, providing individually programmable bias voltages with 0.24 V/LSB resolution, controlled via dedicated 12-bit DACs. The regulator has a nominal 3 mV peak-to-peak ripple and a 100 $\mu$s settling time. Individual bias current monitoring is carried out by means of high-voltage, high-side current sense amplifiers with a 0-5 mA dynamic range.
Regulated voltages, bias currents, and the temperature of the SiPM matrix are sensed via dedicated 12-bit ADC channels, thus completing the slow control chain. An onboard Cortex M4 microprocessor handles all slow control routines and monitors the operational status of each readout channel. All described components are shown in the 3D model of the mezzanine board of Figure~\ref{fig:3Delettronica} top and bottom layers.\\

A total of three SiPM boards and three Mezzanine Boards are currently in production. As an additional R$\&$D step, one of the SiPM boards will be populated with S14160-3010PS~\cite{SiPMdatasheet}, 10 $\mu$m pixel-size SiPMs. This experimental configuration will be tested and compared with the current one for possible future Crilin developments with enhanced timing resolution, better rate capability and improved TNID radiation hardness.

\subsection{Mechanics}\label{Subsec:Subsec4}
\begin{figure}[ht!]
    \begin{center}
        \centering
        \includegraphics[width=.7\textwidth]{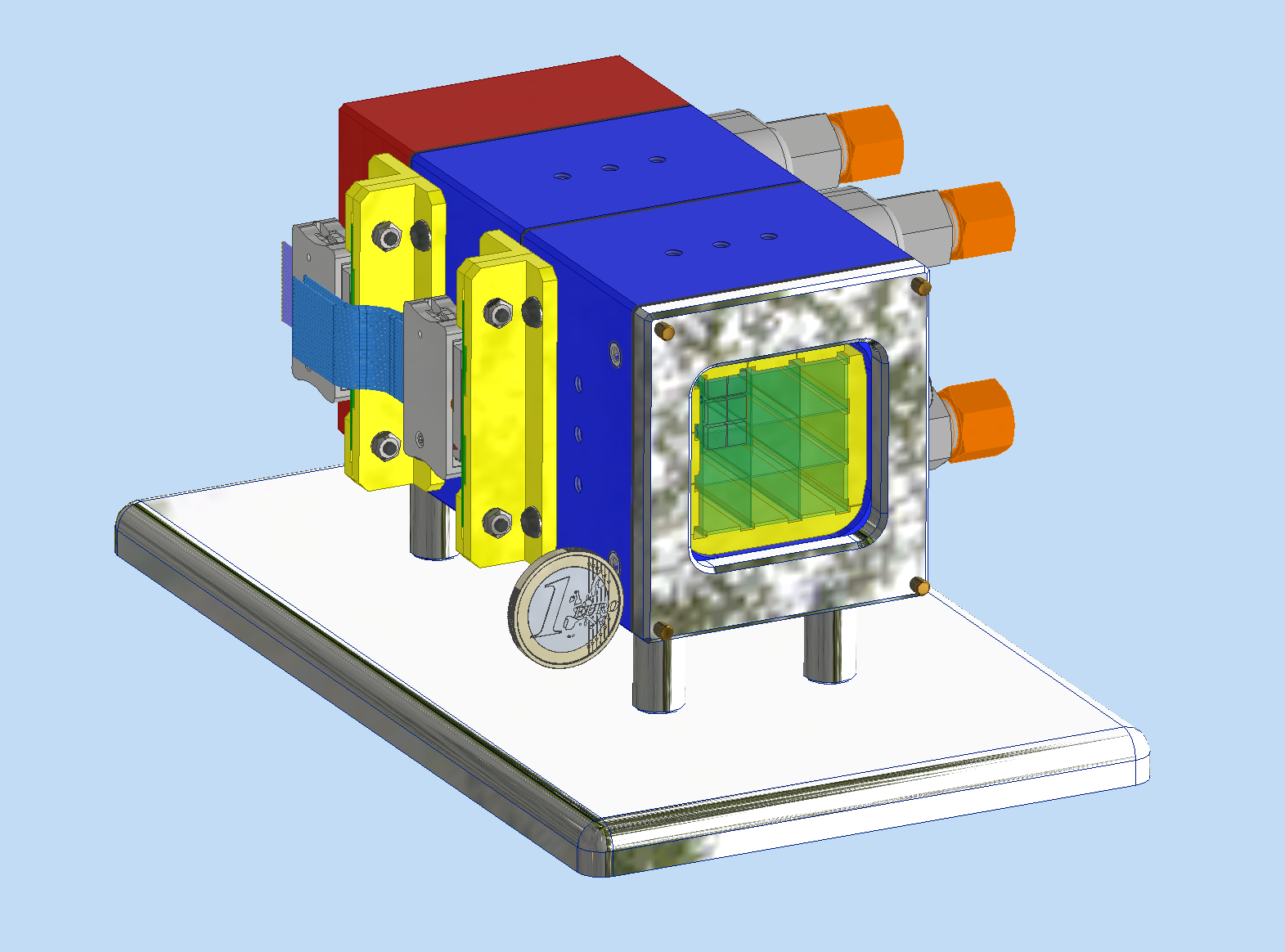}
    \end{center}
    \caption{CAD 3D model of Crilin Prototype (Proto-1).}
    \label{fig:mech1.png}
\end{figure}

In the current design, the prototype consists of two sub-modules, each composed of a 3-by-3 crystal matrix. The sub-modules are arranged in a series and assembled by bolting, obtaining a compact and small calorimeter (Figure~\ref{fig:mech1.png}).

\begin{figure}[ht!]
    \begin{center}
        \centering
        \includegraphics[width=.7\textwidth]{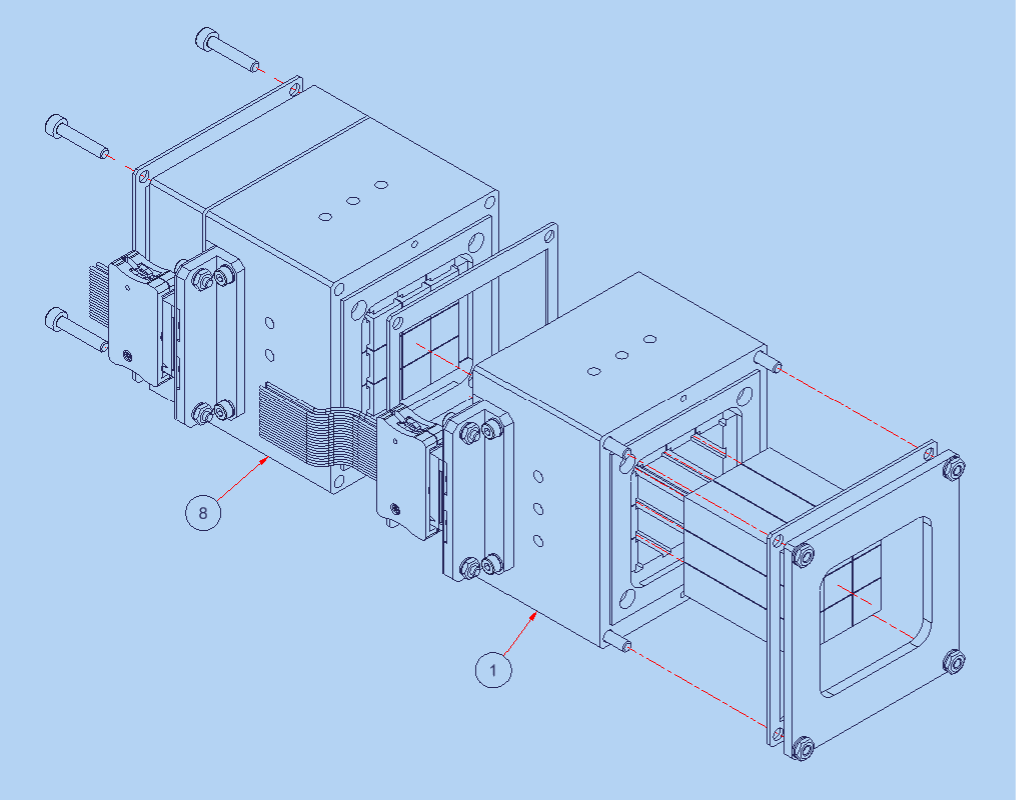}
    \end{center}
    \caption{Cross-section of prototype parts, 3D model.}
    \label{fig:mech2.png}
\end{figure}

Each crystal matrix is housed in a light-tight case which also embeds the front-end electronic boards and the heat exchange needed to cool down the SiPMs. The mechanical architecture of the prototype comprises the following key elements:
\begin{itemize}
    \item The cases, which house each crystal matrix and embed the front-end electronic boards. They are manufactured in common acrylonitrile butadiene styrene (ABS) plastic to minimize the thermal exchange with both the external environment and between the modules.
    \item The locking plates, to which the positioning and blocking of crystals are entrusted, also manufactured in ABS. This solution eases the assembling, positioning, and locking of the crystals matrix. 
    \item The hydraulic connectors, which transport dry gas into the individually sealed modules; the dry gas is circulated inside the active volume of the prototype to prevent condensation.
    \item In between the modules are installed seals, which make each sub-module light-tight. The modules are bolted together using special screws that allow assembling the modules in series. Tedlar{\textregistered} windows close the calorimeter at either end. 
\end{itemize}
The prototype cross section, comprising two modules of 9 crystals each, along with the relative locking plates, is shown in Figure~\ref{fig:mech2.png}.

\subsubsection{The Cooling System}
\begin{figure}[ht!]
    \begin{center}
        \includegraphics[width=0.9\textwidth]{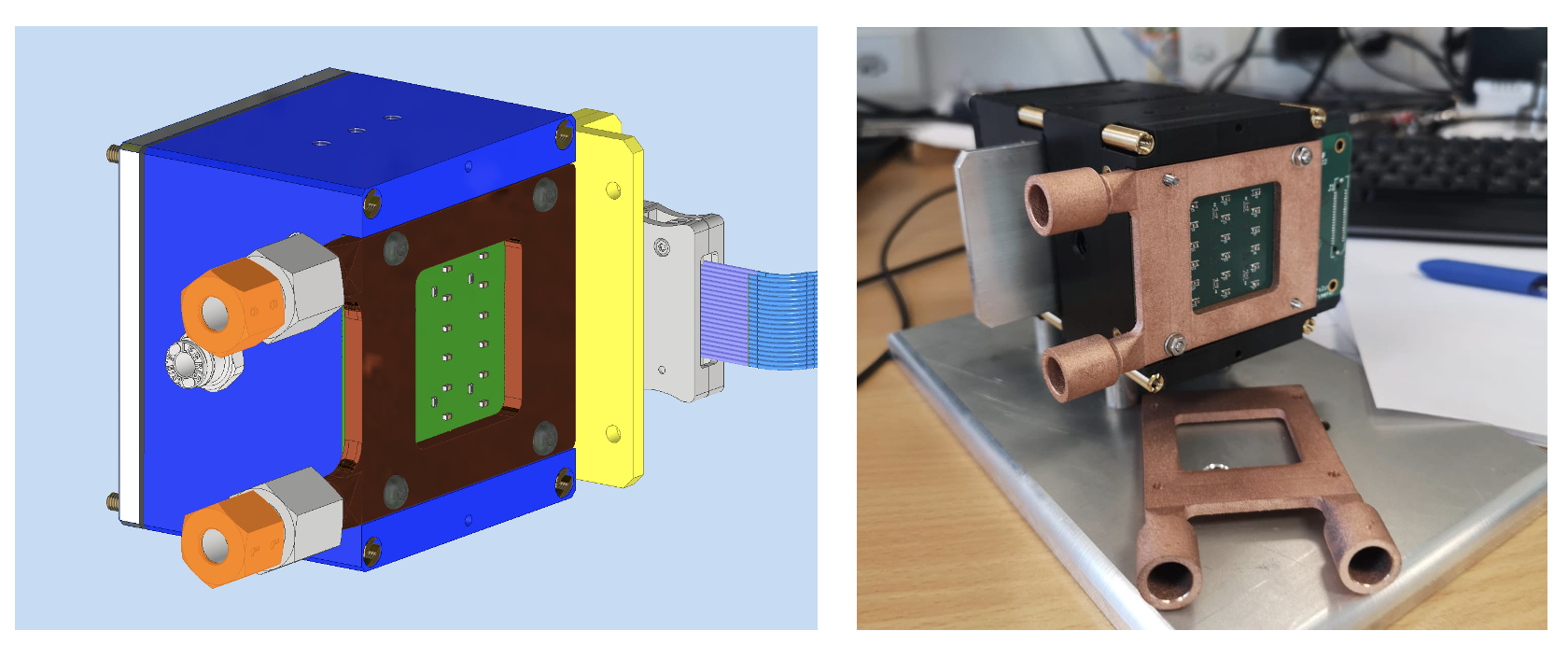}
    \end{center}
    \caption{Cold plate heat exchanger, made of copper, is mounted over the electronic board: 3D Model (left) and assembling on the prototype (right).}
    \label{fig:mech3.png}
\end{figure}

SiPMs must be cooled during operation, so as to improve and stabilise device performance after irradiation; our design is capable of removing the heat load due to the increased photo-sensor leakage current after exposition to a 1-MeV-neq fluence of 10$\mathrm{^{14}\; n_{1MeV}/cm^2}$. The total heat load has been estimated as 350 mW per crystals (per two channels).\\
The cooling system consists of a cooling plant and a cold plate heat exchanger in direct contact with the electronic board. It will provide the optimum operating temperature for the electronics and SiPMs at around 0$^\circ$C. The cooling plant supplies the cold plate with a glycol based water solution at the required flow, temperature and pressure.\\
A cold plate heat exchanger, made of copper, is mounted over the electronic board and the glycol based water solution passes through the deep drilled channels to absorb the heat generated by the SiPMs. To improve the thermal performance of the cold plate, a micro-channel fin structure has been chosen to provide high thermal performance in a compact size (see Figure~\ref{fig:mech3.png}). The cold plate is made by brazing a cover onto the base. The coolant inlet pipe and the outlet pipe are also connected to the cold plate by brazing. The micro-channel fins on the top side of the base part enable the cold plate to cool effectively.

\subsection{First Tests on SiPMs and FEE}\label{Subsec:Subsec5}
\begin{figure}[ht!]
    \centering
     \begin{tabular}{cc} 
    \includegraphics[width=0.47 \textwidth]{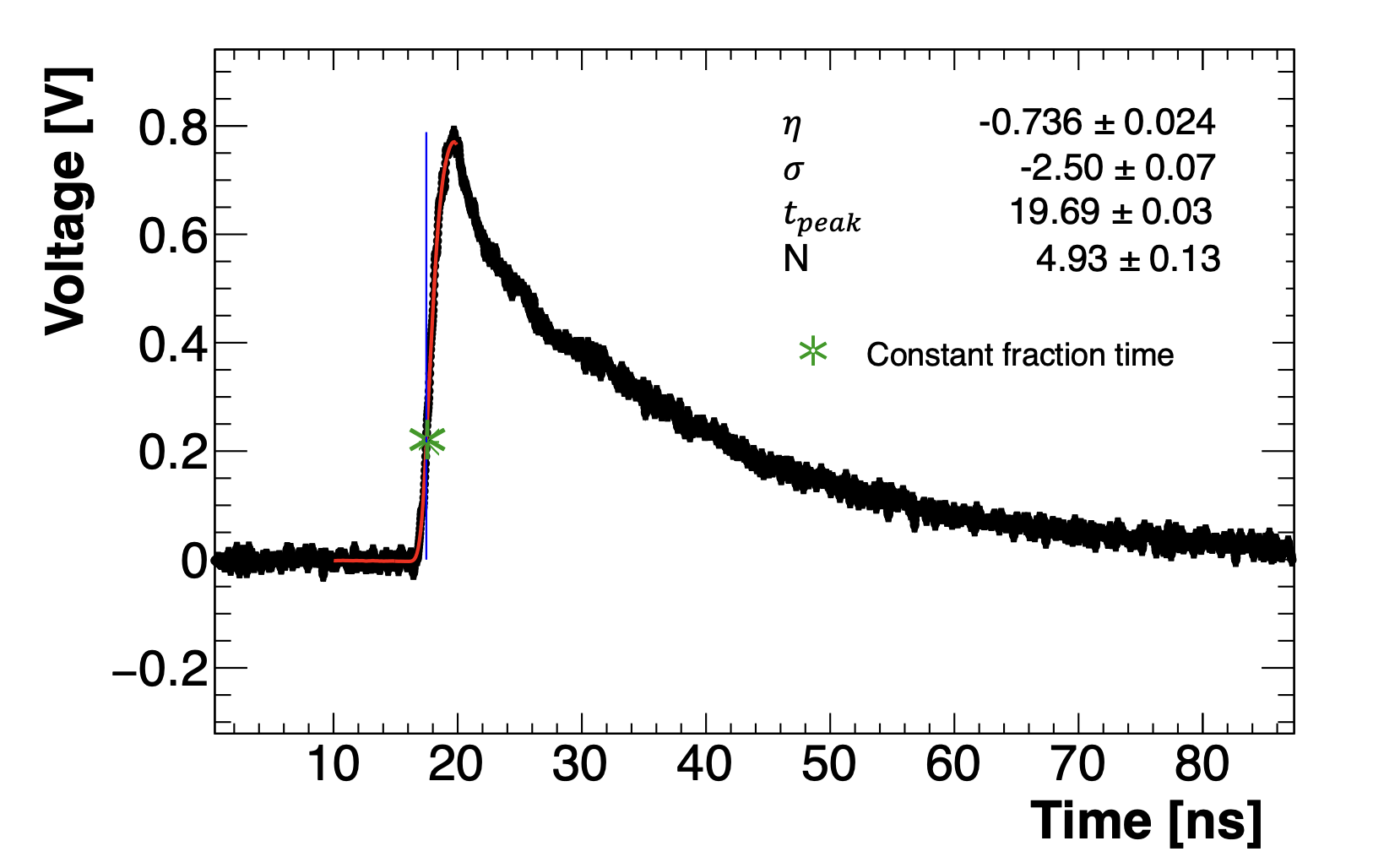} & 
    \includegraphics[width=0.5  \textwidth]{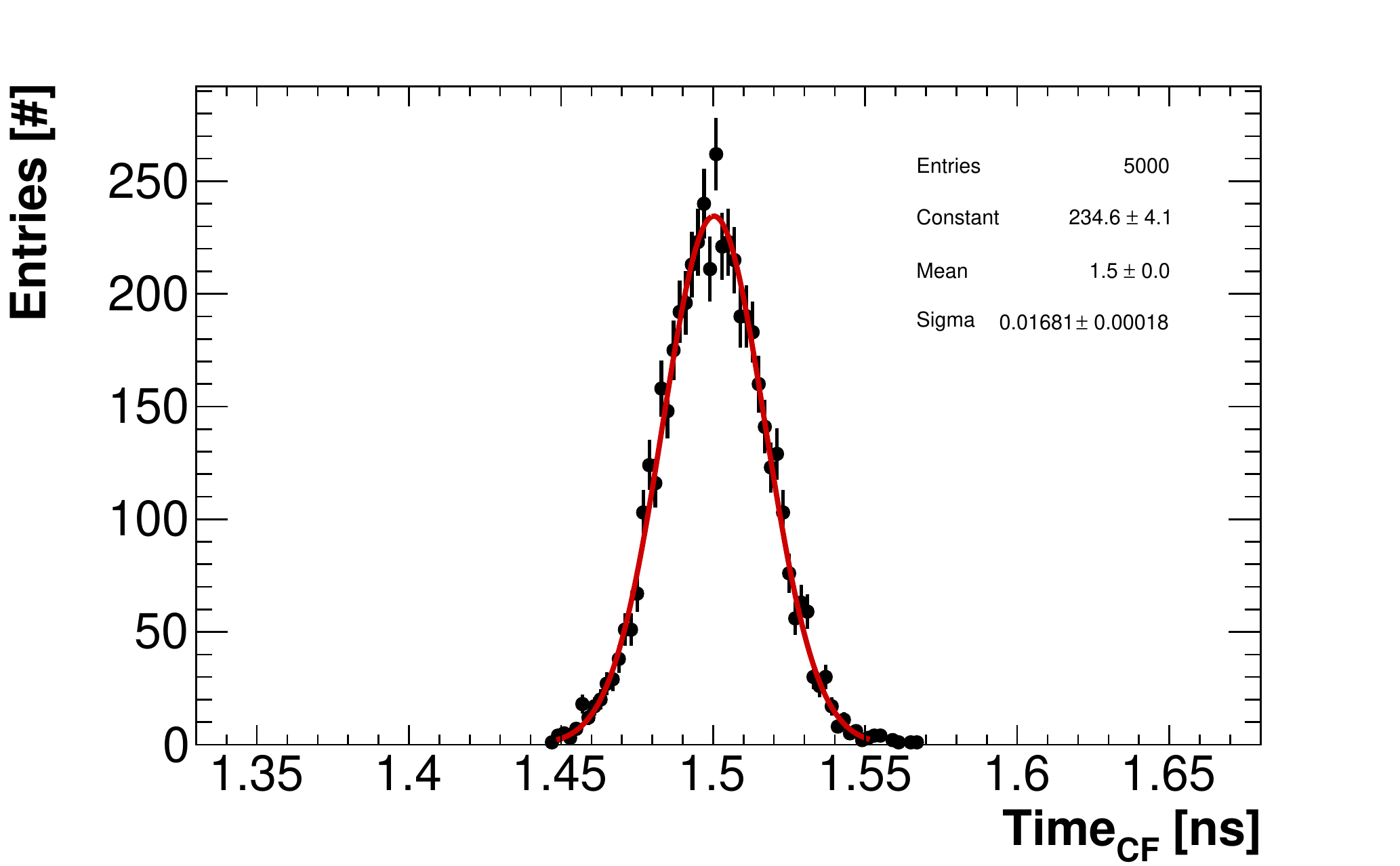} 
    \end{tabular}
    \caption{ Left: SiPM waveform in response to a laser pulse, sampled at 40 GS/s. Log-normal fit on the rising edge is overlayed. Right: reconstructed time distribution for 700 mV amplitude waveforms sampled at 40 Gsps.}
    \label{fig:waveformSiPMs}
\end{figure}

   A first characterisation of the timing performance of the front-end system was carried out using a c10196 picosecond UV laser source by Hamamatsu ~\cite{laser}, by uniformly exposing two S14160-3015PS SiPMs to the laser beam. The photo-detectors were connected in series, and their signals were amplified using the front-end electronics described in section ~\ref{Subsec:Subsec3}, implemented on a prototype board, while an external HV supply was used for biasing. The signals acquisition is made with a SDA820Zi-B Teledyne LeCroy Oscilloscope, and different runs were carried out either by changing the laser pulse amplitude or by changing the oscilloscope sampling rate in the range 2.5 - 40 Gsps. A total of 5000 events was recorded for each run and the acquisition was triggered using the laser's trigger output. SiPM signals collected with this setup (Figure~\ref{fig:waveformSiPMs}-left) show a considerably sharp waveform with a $\sim$2 ns rise time and a $\sim$70 ns full width. 
The data taking phase consisted of three measurement sets:\begin{enumerate}
    \item Constant laser pulse amplitude (adjusted in order to have signals with 1 V peak amplitude) and fixed 40 Gsps sample rate, while laser repetition rate was increased from 50 kHz up to 5 MHz in 5 increments.
    \item Fixed laser amplitude (as before) and fixed 100 kHz laser repetition rate, while the oscilloscope sample rate was swept in the range 2.5 to 40 Gsps. 
    \item Fixed laser repetition rate and sampling frequency (as per points 1 and 2, respectively), while the laser amplitude was swept over the FEE dynamic range in 6 steps.
\end{enumerate}
The laser trigger output was used as a trigger source for all of the measurements: all waveform times hereby reported are thus relative to the laser trigger.\\

\subsubsection{Timing reconstruction method}
\begin{figure}[ht!]
    \centering
     \begin{tabular}{cc} 
    \includegraphics[width=0.49 \textwidth]{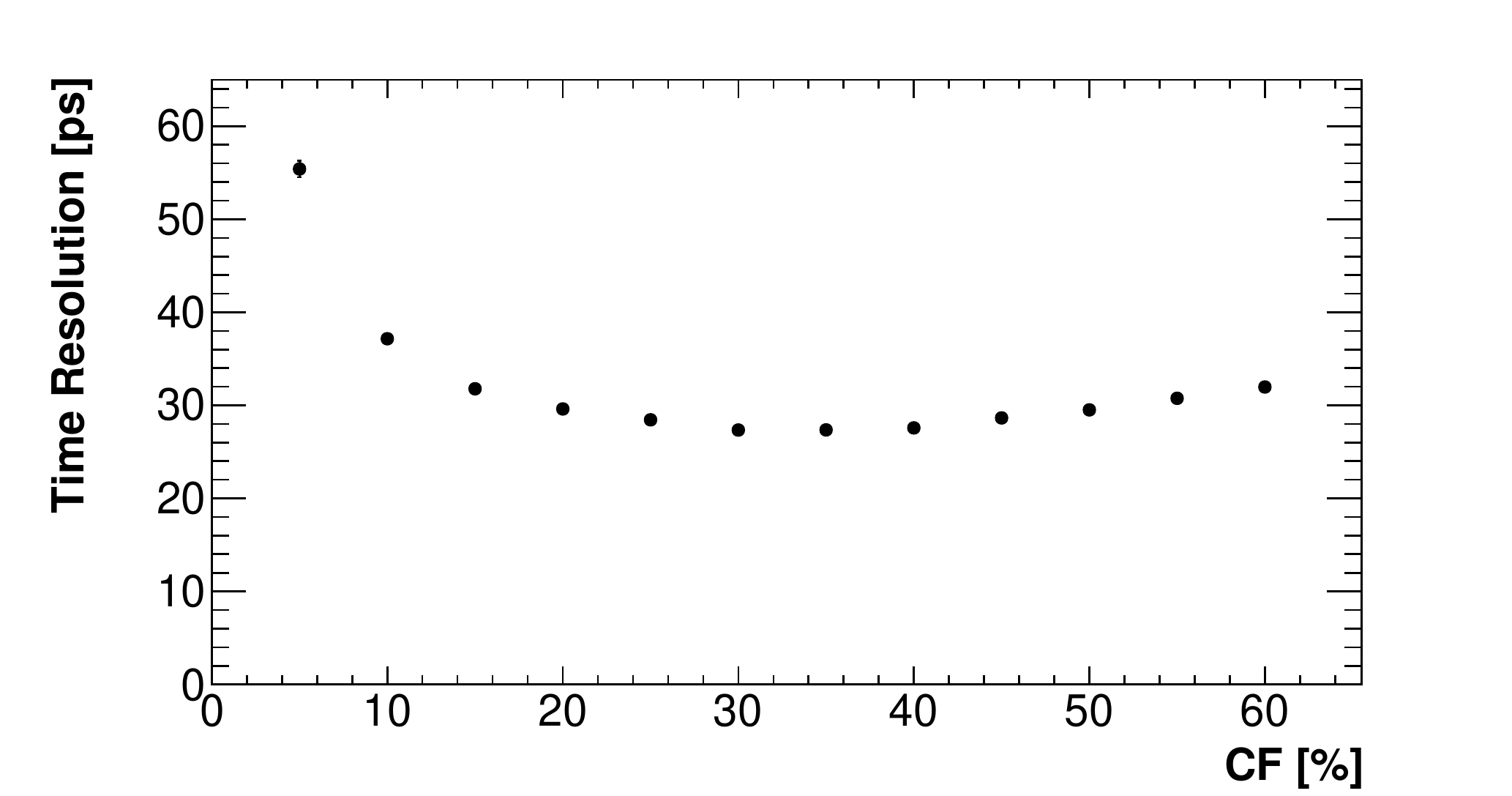} & 
    \includegraphics[width=0.49
    \textwidth]{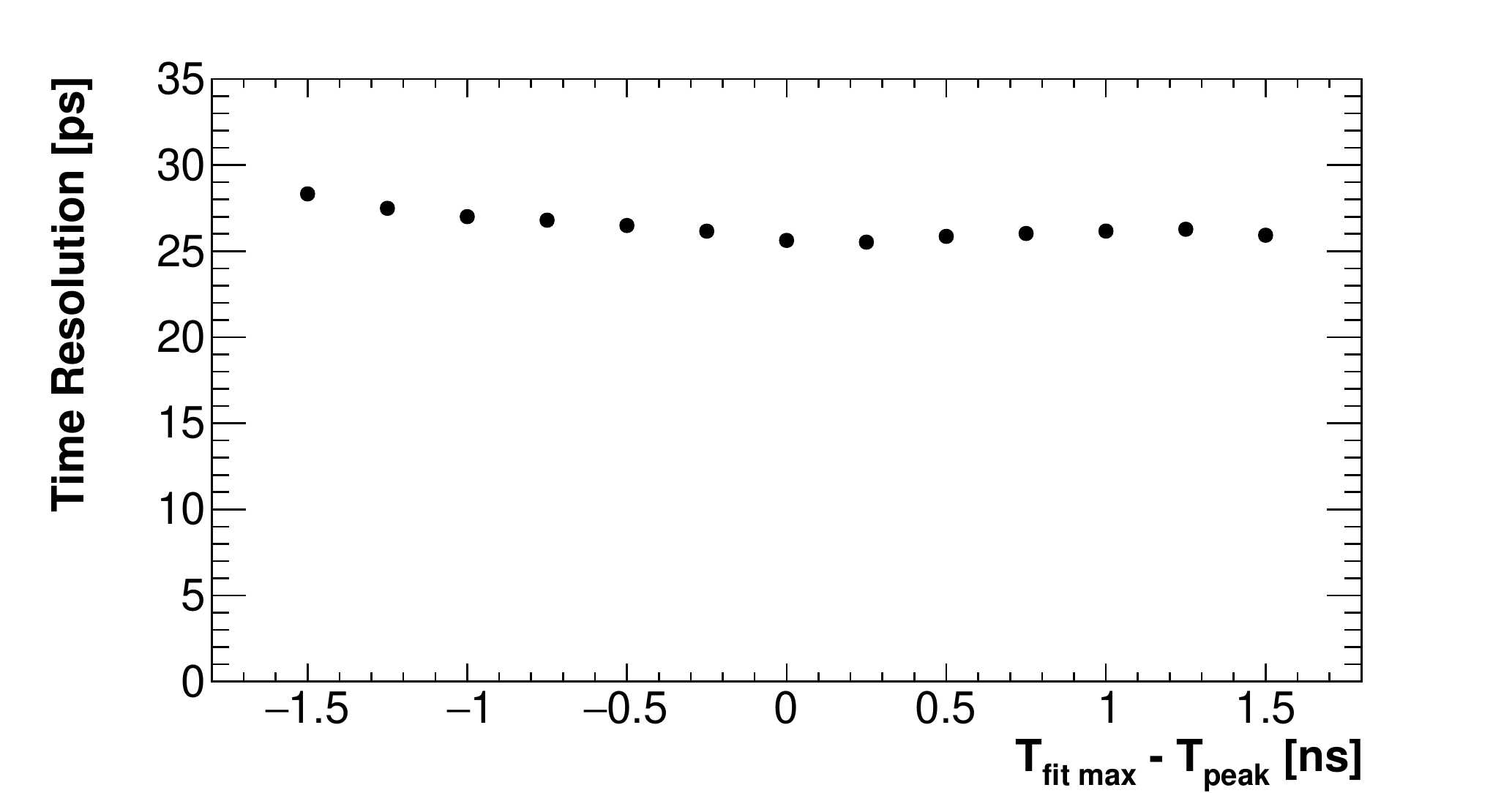} 
    \end{tabular}
    \caption{Optimization of the log-normal fit constant fraction (left) and a fit window upper limit scan (right) by minimisation of the timing resolution.}
    \label{fig:CFandtmax_logn}
\end{figure}

In order to perform the pulse timing reconstruction, a log-normal fit described by Eq.~\ref{logn} was applied to the rising edge of SiPM pulses, as shown in Figure~\ref{fig:waveformSiPMs}-left.

\begin{equation}
\begin{array}{l}
   f(x)=N\frac{\eta}{\sqrt{2\pi}\sigma \sigma_0}\,exp\Bigg\{-\frac{1}{2\sigma_0^2}\,\bigg[log\Big(1-\frac{\eta}{\sigma}(x-t_{peak})\Big)\bigg]^2\Bigg\} , \\
\\
\sigma_0=\frac{2}{2.35}\,log\bigg(\frac{2\eta}{2.35}+\sqrt{1+\Big(\frac{2\eta}{2.35}\Big)^2}\bigg)
 \end{array}    
\label{logn}
\end{equation} 
where N is a normalization factor, $t_{peak}$ is the pulse peak time, $\sigma$ is the FWHM/2.35, while $\eta$ is the asymmetry parameter.
The pulse time was then extracted from the fit using a constant fraction technique. For this purpose, the peak amplitude of each waveform was evaluated by interpolating the waveform peak using a ROOT TSpline object ~\cite{Spline} and finding the local pulse maximum. The timing resolution was then evaluated as the RMS of a Gaussian fit applied to the reconstructed time distributions. For these measurements, the trigger jitter of the laser systems was not taken into account, and represents a contribution included in the timing resolution constant term evaluated in the following sections.\\
The constant fraction value was optimised by minimising the timing resolution, as shown in Figure~\ref{fig:CFandtmax_logn}-left, yielding a best value of 30\% of the peak amplitude. A second optimisation was carried out on the fit range: the lower bound of the fit gate was fixed to $-$12 ns with respect to the waveform peak time, while the upper one was again optimised by minimising the timing resolution. The fit range was thus selected as $[T_{\text{peak}} -12\, \text{ns}, \, T_{\text{peak}} - 0.5\,\text{ns}]$, as shown in Figure \ref{fig:CFandtmax_logn}-right.

\begin{figure}[ht!]
    \centering
     \begin{tabular}{cc}    
    \includegraphics[width=0.49 \textwidth]{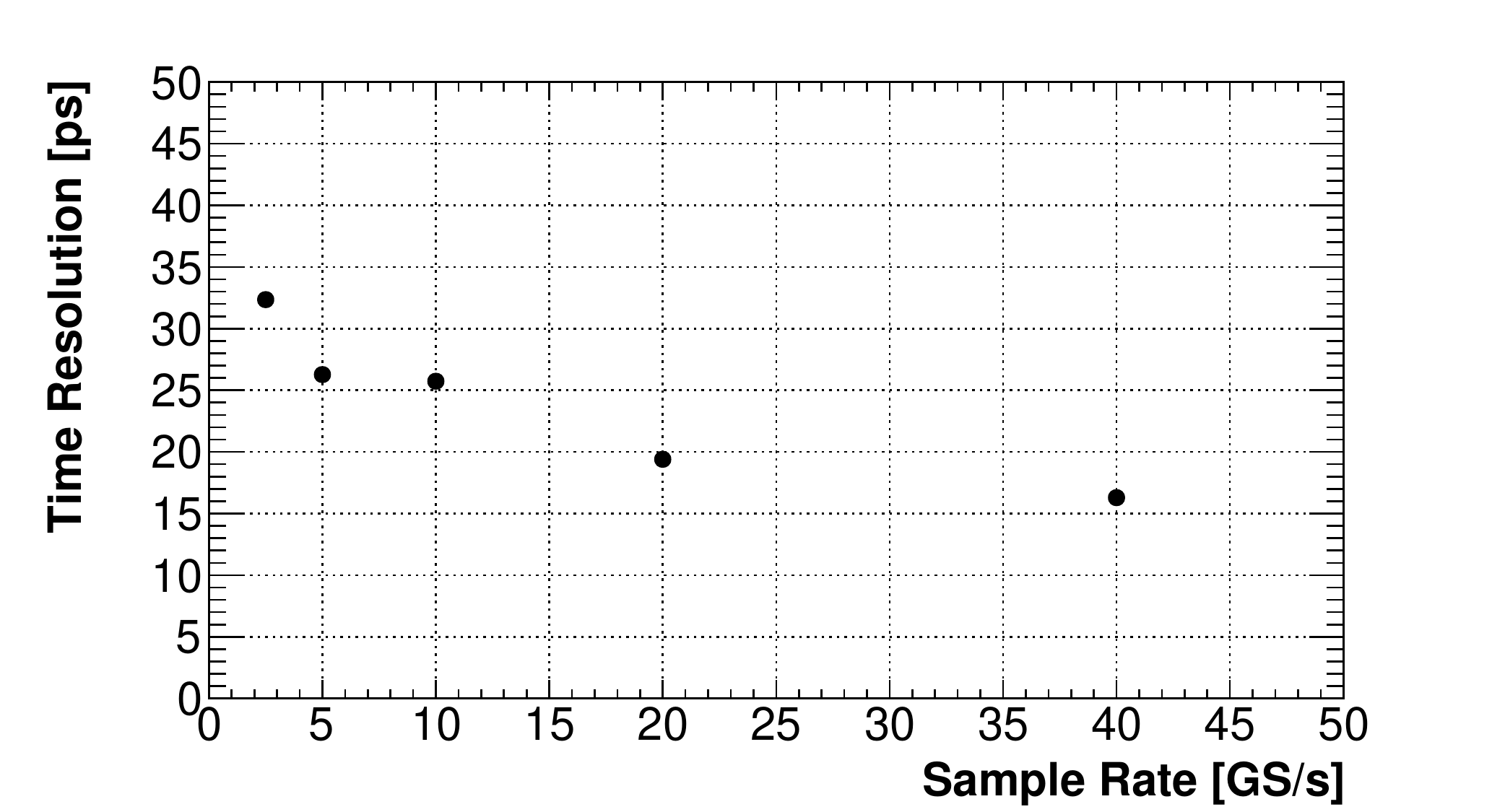} & 
   \includegraphics[width=0.49
   \textwidth]{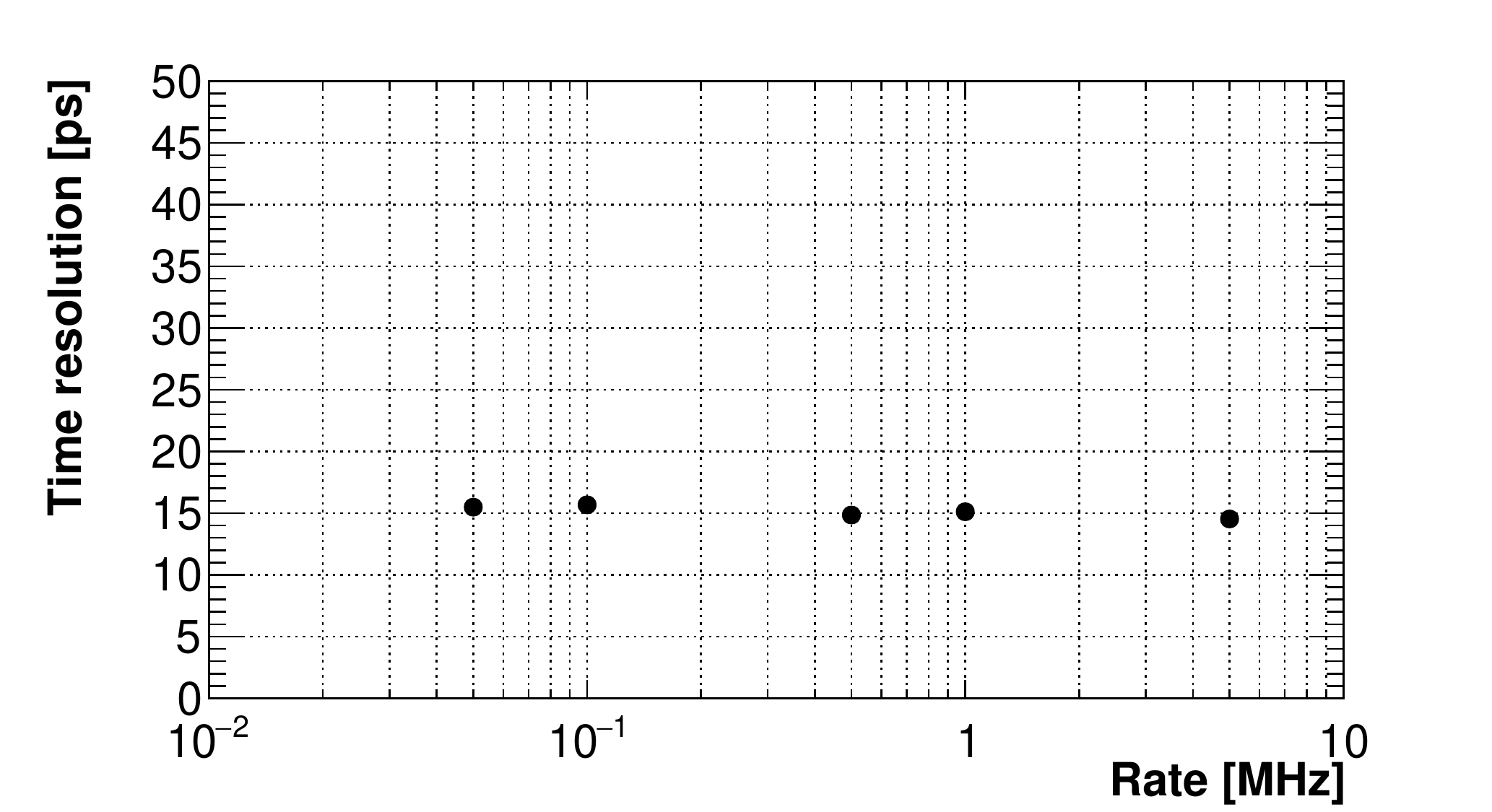} 
    \end{tabular}
    \caption{Effect of sampling frequency (left) and laser repetition rate (right) on timing resolution} 
    \label{fig:TimevsSamplingRate_logn} 
\end{figure}

\paragraph{Effect of laser repetition rate on timing resolution}
In order to asses the effect of the laser repetition rate, a first study was carried out by gradually increasing the laser pulse frequency. For this run, a 40 Gsps sampling rate was used. The laser amplitude was kept constant and adjusted in order to obtain signals with 1 V peak amplitude. The laser repetition rate was then swept from 50 kHz up to 5 MHz in 5 increments. The waveform profiles remained unchanged throughout the test and no effect on the timing resolution was found for laser repetition rates in the range 50 kHz - 5 MHz, as shown in Figure~\ref{fig:TimevsSamplingRate_logn}-right.\\
The purpose of this test was to evaluate the behavior of a single calorimeter readout channel with respect to various possible experimental signal rates. In the case of a Muon Collider, the timing and rate requirements will ensue from the expected luminosity requirements and detector design choices: currently, the designed Muon Collider bunch crossing rate is $\sim$100 kHz at $\sqrt{s}$=1.5 TeV.

\paragraph{Effect of sampling frequency on timing resolution}
A second set of measurements was then performed in order evaluate how the achievable timing resolution was affected by the sampling rate. For this test, the laser amplitude was kept constant (as before) and a fixed 100 kHz laser repetition rate was used, while the oscilloscope sample rate was swept in the range 2.5 to 40 Gsps in 5 steps. The effect of the sampling rate is summarised in Figure~\ref{fig:TimevsSamplingRate_logn}-left, which shows the timing resolution scaling from the worst-case $\sigma_t\sim32$ ps obtained at 2.5 GS/s to $\sigma_t\sim 15$ ps at 40 GS/s. 

An additional data-set was acquired using a 1 GS/s sampling rate. For this run, however, an insufficient number of sampled points (2 or 3) are available on the rising edge of the waveform for the log-normal fit to achieve an effective implementation of our timing reconstruction method, at least without any additional corrections and studies which are beyond the current scope.

\begin{figure}[ht!]
    \centering
     \begin{tabular}{cc}    
    \includegraphics[width=0.46 \textwidth]{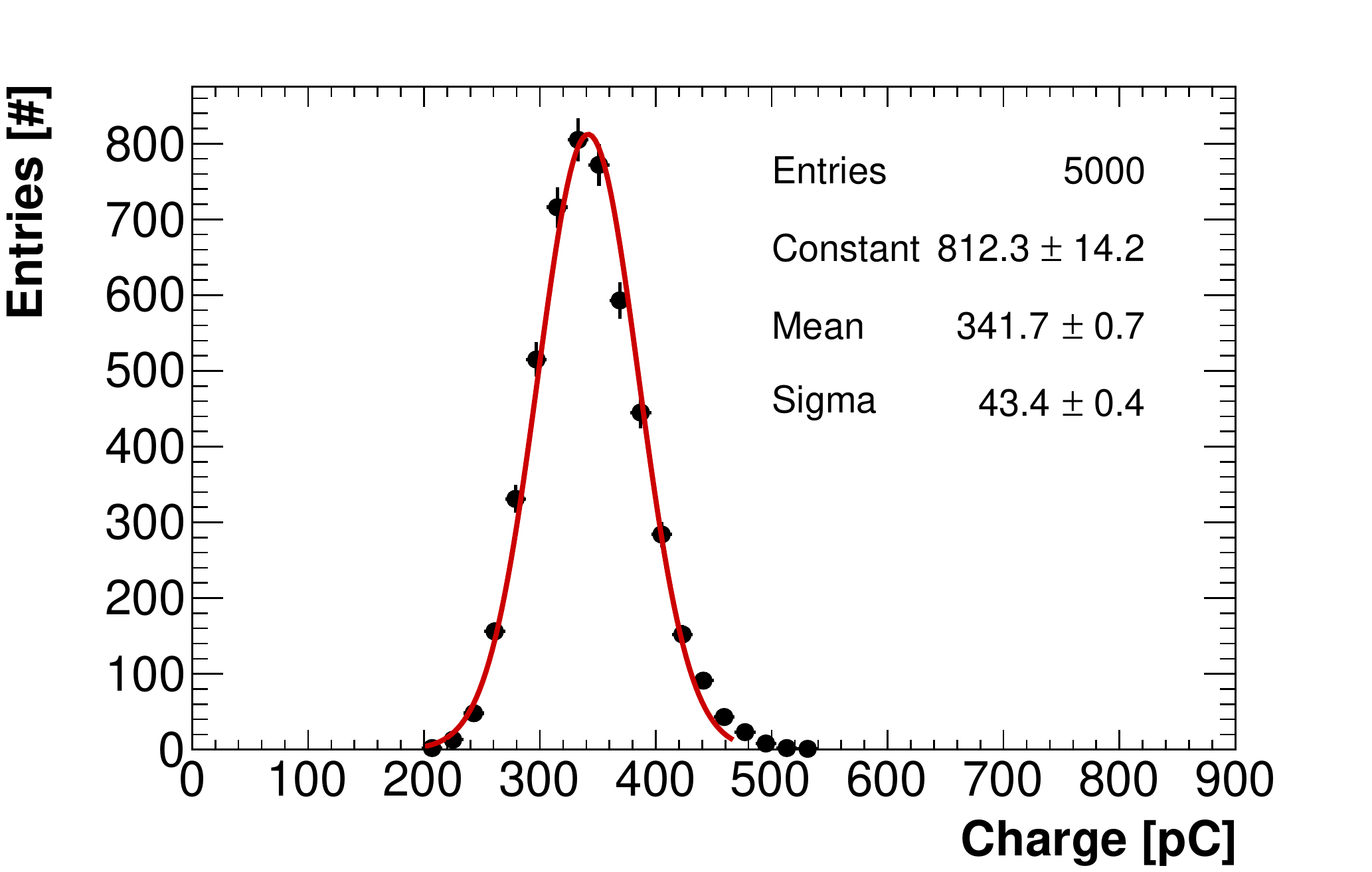} & 
   \includegraphics[width=0.54
   \textwidth]{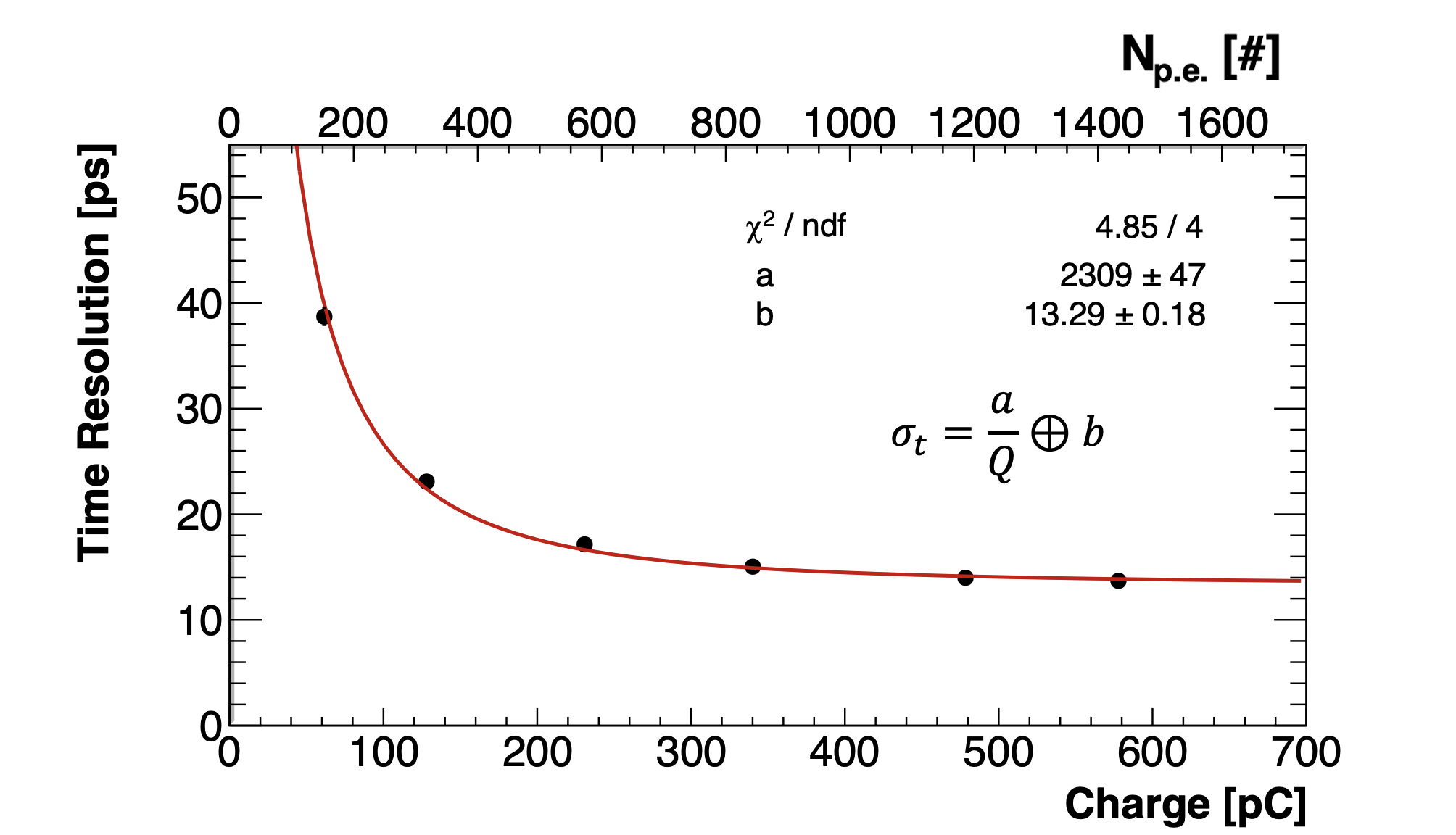} 
    \end{tabular}
    \caption{ Example of charge distribution for 1V peak amplitude waveforms (left). Time resolution as a function of charge and N$_{\text{p.e.}}$ (right).
    } 
    \label{fig:TimevsChargePE_logn}
\end{figure}

\paragraph{Timing resolution dependence on charge.}
The charge dependence of the time resolution was evaluated with a dedicated set of measurements carried out at 40 GS/s, using fixed laser repetition rate of 100 kHz, and six different laser amplitude settings. 
Pulse charges were evaluated by integrating each waveform in a [T$_{\text{peak}}$-20 ns; T$_{\text{peak}}$+60 ns] gate, taking into account the 50$\,\Omega$ oscilloscope input impedance, and with T$_\text{{peak}}$ the waveform peak time. For each run, the charges showed a Gaussian distribution, as reported in Figure~\ref{fig:TimevsChargePE_logn} (left), and the charge reference value for each run was evaluated as the mean value of a normal fit.\\
The charge dependence of the timing resolution is summarised in Figure \ref{fig:TimevsChargePE_logn} (right). Data were fitted using the model of Eq.~\ref{s_t e Q}  ~\cite{Time e p.e.}:
\begin{equation}
\sigma_t=\frac{a}{Q}\oplus b
\label{s_t e Q}
\end{equation}
where $a$ accounts for the effects of photo-statistics and of the system rise time, while $b$ is the constant term representing the asymptotic limit on the time resolution.\\
The mean photo-electron numbers obtained with these runs was derived as 
 \begin{equation}
     N_{p.e.}= \frac{Q}{G_{FEE}\,G_{SiPMs}\,e} 
 \end{equation}
 
 where  $G_{FEE} = 7 $ is the FEE gain,  $G_{SiPMs}$ = 3.6$\times$10$^5$ ~\cite{SiPMdatasheet} is the SiPM gain and $e$ is the electron charge. As a result, a conversion factor of $\sim$2.48 photo-electrons per pC was calculated, as shown in Figure~\ref{fig:TimevsChargePE_logn}-right. The maximum charge value measured in Figure~\ref{fig:TimevsChargePE_logn}-right ($\sim$580 pC) corresponds thus to $\sim$1440 photo-electrons, for SiPM peak pulse amplitudes of $\sim$1600 mV, close to the full-scale FEE dynamic range.\\ 

This charge scan demonstrates the high performance that the calorimeter electronics can reach by showing a time resolution that is already less than 40 ps even at low charges (50 pC - 124 photo-electrons) and an impressive constant term $b$ of $\sim$ 13 ps.

\section{Conclusion}
The Crilin design aims to overcome the classic rigid distinction between homogeneous and sampling calorimeters, trying to establish a good compromise between the two in order to optimize the requirements in view of a future Muon Collider.\\
The modern quest for high intensity experiments, indeed, makes it necessary to develop detectors capable of supporting innovative reconstruction techniques, enabling superior signal extraction from harsh and high-rate backgrounds. Although very high granularity solutions with millions of readout channels enabling particle-flow techniques (seen for example with the CALICE W-Si ECAL) are able to cope with the high-rate beam-induced background which could be encountered, for example, in a future Muon Collider experiment, chief disadvantages of this solution are technological complexity and cost. Furthermore, in this configuration, the energy resolution will be intrinsically limited by the sampling configuration.
Our solution, a semi-homogeneous calorimeter with longitudinal segmentation and superior timing resolution (less then 100 ps for each individual readout channel), allows to overcome these issues, while enabling superior event reconstruction strategies, thanks to the use of 5D discrimination.  Enhanced clustering techniques could thus be implemented on this platform, for the efficient rejection of virtually any type of background. Furthermore, the semi-homogenous longitudinal segmentation can be easily tuned, both in simulation and in the real detector, to better match different experimental conditions (in terms of energy scale, resolution, and the expected background spectrum) with a high degree of reconfigurability and minimal optimisation effort.\\
Before the construction of the prototype, the individual components were evaluated. In particular: 
\begin{itemize}
    \item irradiation studies of crystals and SiPMs;
    \item a preliminary two crystals test beam at BTF with 500 MeV in July 2021 and at CERN in August 2021.
\end{itemize}

\noindent A Crilin prototype, composed of two layers of nine crystals each and operating at 0~\temp, will be built during 2022. Our goal is to test its performance with 500 MeV electrons at BTF and with a high energy beam ($>$ 100 GeV) at CERN at end of July 2022.

\section{Acknowledgements}
This work was developed within the framework of the International Muon Collider Collaboration (https://muoncollider.web.cern.ch), where the Physics and Detector Group aims to evaluate potential detector R$\&$D to optimize experiment design in the multi-TeV energy regime. This project has received funding from the European Union’s Horizon 2020 Research and Innovation program under GA no 101004761.\\
The authors wish to thank the LNF Division Research and SPCM departments for their technical and logistic support and are grateful to the people in Enea Casaccia and Enea FNG facilities for their extensive collaboration during the irradiation campaign of crystals and photo-sensors.	

%\section*{References}


\begin{thebibliography}{9}
%----------------------Introduzione

\bibitem{map} Muon Accelerator Program, \url{https://map.fnal.gov/}
\bibitem{nozzle} N.V. Mokhov and S.I. Striganov, Phys. Procedia 37 (2012) 2015.
\bibitem{bib} N.V. Mokhov and C.C. James, Fermilab-FN-1058-APC (2018).
\bibitem{XX1} N. Bartosik et al., Detector and Physics Performance at a Muon Collider, JINST 15 P05001 (2020).
\bibitem{Calice} L. Linssen et al., Physics and Detectors at CLIC: CLIC Conceptual Design Report, CERN Yellow Report CERN-2012-003.
\bibitem{snowmass} N. Bartosik et al., Simulated Detector Performance at the Muon Collider, arXiv:2203.07964v1.
\bibitem{aps} L. Sestini \emph{et al}, Jet Reconstruction performance at Muon Collider with Beam-induced Background, APS April Meeting 2021.
\bibitem{ACTS} X. Ai et al., A Common Tracking Software Project, arXiv:2106.13593v1.
\bibitem{Pandora} M.A. Thomson, Particle flow calorimetry and the PandoraPFA algorithm, Nuclear Instruments and Methods in Physics Research Section A, Volume 611, Issue 1, 2009, Pages 25-40.
\bibitem{KT} S. D. Ellis et al., Successive Combination Jet Algorithm For Hadron Collisions, arXiv:9305266v1.
%----------------------Camilla
\bibitem{fluka} A. Ferrari et al., FLUKA: a multi-particle transport code, CERN-2005-10 (2005), INFN/TC$\_$05/11, SLAC-R-773.
\bibitem{Collamati:2021sbv} F. Collamati et al., Advanced assessment of beam-induced background at a Muon Collider, JINST 16 (2021) P11009.
%---------------------- REF section3 ---------------------- 
\bibitem{SIC} Siccas official page official page, \url{http://www.siccas.com/PbF2LeadfluorideCrystal.htm}
\bibitem{articolo_irraggiamenti} Cemmi, A., et al. Radiation study of Lead Fluoride crystals. Journal of Instrumentation 17.05 (2022): T05015.

\bibitem{Calliope}Calliope Brochure, \url{https://www.enea.it/it/seguici/pubblicazioni/pdf-opuscoli/calliope.pdf}
\bibitem{zhu1996study} Zhu, R-Y et al., A study on the properties of lead tungstate crystals, Nuclear Instruments and Methods in Physics Research Section A: Accelerators, Spectrometers, Detectors and Associated Equipment 376.3 (1996): 319-334.
\bibitem{ren2001optical}G.H. Ren et al., Optical Absorption on Cubic $\beta$-PbF$_{2}$ Crystals. Chinese Physics Letters 18.7 (2001): 976-978.
\bibitem{achenbach1998radiation} P. Achenbach et al., Radiation resistance and optical properties of lead fluoride Cherenkov crystals. Nuclear Instruments and Methods in Physics Research Section A: Accelerators, Spectrometers, Detectors and Associated Equipment 416.2-3 (1998): 357-363.
\bibitem{kozma2002radiation} P. Kozma,  et al., Radiation resistivity of PbF$_{2}$ crystals. Nuclear Instruments and Methods in Physics Research Section A: Accelerators, Spectrometers, Detectors and Associated Equipment 484.1-3 (2002): 149-152.
\bibitem{FNG} FNG website, \url{http://www.fusione.enea.it/LABORATORIES/Tec/FNG.html.it}
\bibitem{mcstas} McStas website, \url{http://www.mcstas.org/}
%---------Prototipo
\bibitem{MATT} AIDAinnova 1st Annual Meeting, \url{https://indico.cern.ch/event/1104064/contributions/4801240/attachments/2417197/4136687/Moulson_220329WP8.pdf}
\bibitem{SiPMdatasheet} Hamamatsu SiPMs datasheet,  \url{https://www.hamamatsu.com/content/dam/hamamatsu-photonics/sites/documents/99_SALES_LIBRARY/ssd/s14160-1310ps_etc_kapd1070e.pdf}
\bibitem{SAMTEC} Samtec cables datasheet, \url{https://www.samtec.com/products/ercd-040-40.00-ted-teu-1-d}

%---------Test laser-------------
\bibitem{laser} Hamamatsu laser source datasheet,
\url{https://www.hamamatsu.com/content/dam/hamamatsu-photonics/sites/documents/99_SALES_LIBRARY/sys/SOCS0003E_PLP-10.pdf}

%\bibitem{Oscilloscopio} Teledyne LeCroy Oscilloscope datasheet, \url{https://cdn.teledynelecroy.com/files/pdf/wavemaster-8zi-b-datasheet.pdf}

\bibitem{Spline} Root Cern TSpline3 class reference, \url{https://root.cern.ch/doc/master/classTSpline3.html}
\bibitem{Time e p.e.} Baldini et al., A cryogenic facility for testing the PMTs of the MEG liquid xenon calorimeter. Nuclear Instruments \& Methods in Physics Research Section A: Accelerators, Spectrometers, Detectors and Associated Equipment, 566 (2006): 294-301.  
%------fine laser-----

%\bibitem{map} Muon Accelerator Program, https://map.fnal.gov/ 
%\bibitem{nozzle} N.V. Mokhov and S.I.Striganov, Phys. Procedia 37 (2012) 2015 
%\bibitem{bib} N.V. Mokhov and C.C. James, Fermilab-FN-1058-APC (2018)
%\bibitem{XX1} N. Bartosik et al., Detector and Physics Performance at a Muon Collider, JINST 15 P05001 (2020).
%\bibitem{ilcsoft} https://github.com/iLCSoft
%\bibitem{clic} CLIC collaboration, CERN-2012-003
%\bibitem{aps} L. Sestini \emph{et al}, Jet Reconstruction performance at Muon Collider with Beam-induced Background, APS April Meeting 2021
%\bibitem{cinque}Brochure Calliope, \url{https://www.enea.it/it/seguici/pubblicazioni/pdf-opuscoli/calliope.pdf}
%\bibitem{sei}FNG website, \url{http://www.fusione.enea.it/LABORATORIES/Tec/FNG.html.it}
%\bibitem{uno} P. Kozma et al., Radiation resistivity of PbF2 crystals, Nucl.Instrum.Meth.A 484 (2002) 149-152.
%\bibitem{BTF} B. Buonomo et al., The Frascati LINAC Beam-Test Facility (BTF) Performance and Upgrades, IBIC2016 (2017) 395-398.

\end{thebibliography}
\end{document}